\documentclass[aps,prd,twocolumn,nofootinbib,superscriptaddress,notitlepage]{revtex4-2}
\usepackage{bm}
\usepackage{amsmath,amsthm,amssymb,mathtools}
\usepackage{verbatim}
\usepackage{color}
\usepackage{tabularx}
\usepackage{tensor}
\usepackage{braket}

\renewcommand{\d}{\textrm{d}}
\def\Lagr{\mathcal{L}}

\def\be{\begin{equation}}
\def\ee{\end{equation}}

\begin{document}

\title{The Role of the Schwinger Effect in Superradiant Axion Lasers}
\author{Bradley Shapiro}
\email{bradley.e.shapiro.gr@dartmouth.edu}
\affiliation{Department of Physics and Astronomy, Dartmouth College, Hanover, NH 03755}
\date{\today}

\begin{abstract}
Superradiance can cause the axion cloud around a rotating black hole to reach extremely high densities, and the decay of these axions can produce a powerful laser. The electric field of these lasers is strong enough that the Schwinger effect may become significant, resulting in the production of an electron-positron plasma. We explore the dynamics between axion lasers and this electron-positron plasma. While there are several mechanisms by which the inclusion of a plasma can impact the laser’s behavior, the most significant of these mechanisms is that the electron-positron plasma imparts an effective mass on the photon. As the plasma frequency increases, axion decay becomes energetically unfavorable, up to the point where the axion no longer decays into photons, shutting off the laser. We find that the impact of the electron-positron plasma on the dynamics of the system depend heavily on the parameters, specifically the axion mass $m_\phi$ and the superradiant coupling $\alpha$, and that we may divide parameter space into three regimes: the unenhanced, enhanced, and unstable regimes. In the unenhanced and enhanced regime, the system will eventually settle into an equilibrium state, emitting a laser of constant luminosity while the number of axions remains constant. In the unenhanced regime, this equilibrium state can be calculated while neglecting the effects of Schwinger production; in the enhanced regime, the equilibrium luminosity is slightly larger than what it would be without Schwinger production. In the unstable regime, the electron-positron plasma suppresses axion decay to the point where the system is never able to reach equilibrium; instead, the axions continue to grow superradiantly. In all three cases, the production of superradiant axions will eventually cause the black hole to spin down to the point where superradiance ceases.
\end{abstract}

\maketitle

\section{Introduction}

The axion has emerged as a leading candidate for dark matter \cite{PhysRevLett.38.1440,PhysRevLett.40.223,PhysRevLett.40.279,Preskill:1982cy,Abbott:1982af,Dine:1982ah,Adams:2022pbo}. A topic of significant research interest is the idea that axions might be detected around rotating black holes \cite{Arvanitaki_2010,Arvanitaki_2011,Brito_2015,Arvanitaki_2017,Davoudiasl_2019}, via superradiance, which allows for the extraction of energy and angular momentum from a rotating black hole \cite{Detweiler:1980uk,Dolan_2007,Brito_2020}. Superradiance is often understood through the concept of a gravitational atom \cite{Arvanitaki_2015,Baumann_2019}, as, in Kerr spacetime, the Klein-Gordon equation permits quasi-bound state solutions, characterized by integer quantum numbers $(n,\ell,m)$, much like those around a hydrogen atom. For a massive scalar field, these quasi-bound states have energy \cite{Brito_2020}
\be E_{n,\ell}=\textrm{Re}(\omega_{n,\ell})=m_\phi\left(1-\frac{\alpha^2}{2(n+\ell+1)^2}\right), \ee
where $m_\phi$ is the scalar mass and the coupling parameter is
\be \alpha=\frac{G m_\phi M_{BH}}{\hbar c}=.037\left(\frac{m_\phi}{10^{-5}\textrm{ eV}}\right)\left(\frac{M_{BH}}{10^{24}\textrm{ kg}}\right). \ee
Importantly, unlike in hydrogen atoms, where the frequency is entirely real, the quasi-bound states around rotating black holes may have a complex frequency, $\textrm{Im}(\omega_{n,\ell})\neq0$. Depending on the sign of this imaginary frequency, the state will experience either exponential growth or exponential decay proportional to $e^{\textrm{Im}(\omega_{n,\ell})t}$. States with exponential growth are referred to as superradiant.

Conceptually, there are a number of ways to understand superradiance. \cite{Bekenstein_1998} showed that, when a wave is incident on a rotating body, the second law of thermodynamics requires, under certain conditions, that the wave be reflected back with greater energy and angular momentum than it initially had. From a particle perspective, superradiance may be understood as a consequence of the Penrose process \cite{Penrose:1969pc,Penrose:1971uk}, whereby an object passing through the ergosphere of a rotating black hole may extract energy and angular momentum from the black hole. In either picture, this energy extraction can only occur through the creation of new particles, because the quasi-bound nature of the aforementioned states means that individual particles cannot become more energetic. Consequently, superradiance increases the population of particles around the black hole. The aforementioned descriptions are classical in nature, i.e. the scalar field is not quantized, which is how superradiance is usually treated. However, there do exist some treatments of superradiance as a quantum phenomenon \cite{Alicki_2018,Balakumar_2020}. These offer some additional insight into the exact mechanism in which superradiant particles are created around a black hole.

In models that include an axion, rotating black holes are therefore expected to produce a dense axion cloud via superradiance. It has been proposed \cite{Rosa_2018,Sen_2018,Bo_kovi__2019,Ikeda_2019,spieksma2023superradianceaxioniccouplingsplasma} that, in certain regions of parameter space, these axions can undergo stimulated decay into photons, creating a powerful laser, referred to in \cite{Rosa_2018} as a BLAST (\emph{Black hole Lasers powered by Axion SuperradianT instabilities}). It has been observed \cite{Rosa_2018,Bo_kovi__2019} that these BLASTs are powerful enough that the Schwinger effect may become significant. The Schwinger effect refers to a quantum mechanical, nonperturbative process whereby a strong electromagnetic field produces electron-positron pairs. Plasma effects are significant in the dynamic of BLASTs \cite{Sen_2018,Bo_kovi__2019,spieksma2023superradianceaxioniccouplingsplasma}, primarily because the plasma imparts photons with an effective mass, which eventually makes axion decay energetically impossible, thereby shutting off the laser. \cite{Rosa_2018} speculated that plasma effects would result in the axion laser shutting off and then restarting in a periodic fashion, although they did not explore the underlying dynamics in detail. \cite{Bo_kovi__2019,Ikeda_2019} performed numerical simulations that seemed to validate the prediction of periodic bursts, whereas the simulations in \cite{spieksma2023superradianceaxioniccouplingsplasma} found that laser-like emission occurs not in bursts, but smoothly; however, all of these simulations assumed an electron-ion plasma acquired via accretion, as opposed to an electron-positron plasma acquired via the Schwinger effect. This paper therefore aims to explore the role of Schwinger production, specifically, in axion lasers.

This paper is structured as follows. In section 2, we review how BLASTs behave in the absence of the Schwinger effect. In section 3, we explore how an electron-positron plasma evolves in the superradiant axion cloud around a rotating black hole, and how that plasma affects the behavior of the photons and axions in that cloud. In section 4, we analyze the resulting Boltzmann equations. Section 5 is the conclusion.

\section{BLASTs without Schwinger production}
We begin by briefly summarizing the work of \cite{Rosa_2018}.\footnote{Note that some numerical values differ from those given in \cite{Rosa_2018}, because we parameterize quantities in terms of $\frac{\alpha}{.037}$, whereas \cite{Rosa_2018} parameterizes quantities in terms of $\frac{\alpha}{.03}$.} Importantly, this approach assumes a homogenous axion cloud in which decay happens uniformly; however, detailed simulations show that the substructure of the axion cloud does indeed play a role in the nature of the emitted lasers \cite{spieksma2023superradianceaxioniccouplingsplasma}. For example, lasers will include a beating pattern that is not predicted under the approach described herein, which results from photons produced inside the axion cloud traveling outward and interacting with the outer parts of the cloud. The simulations needed to predict such phenomena are beyond the scope of this paper, and we restrict ourselves to a simpler, analytic approach, which should be understood as a simplified model of a complicated system.

For scalar fields, the fastest-growing state is the $2p$-state ($n=2$, $\ell=m=1$). The $2p$-state is approximately toroidal in shape, with the following dimensions:
\begin{widetext}
\begin{align}
\textrm{Major radius:}&\hspace{1.5cm}\langle r\rangle=\frac{5\hbar}{cm_\phi\alpha}=2.6\left(\frac{\alpha}{0.037}\right)^{-1}\left(\frac{m_\phi}{10^{-5}\textrm{ eV}}\right)^{-1}\textrm{ m} \\
\textrm{Minor radius:}&\hspace{1.5cm}\Delta r=\frac{\sqrt{5}\hbar}{cm_\phi\alpha}=1.2\left(\frac{\alpha}{0.037}\right)^{-1}\left(\frac{m_\phi}{10^{-5}\textrm{ eV}}\right)^{-1}\textrm{ m} \\
\textrm{Volume:}&\hspace{1.5cm}V=2\pi^2\langle r\rangle\Delta r^2=71\left(\frac{\alpha}{0.037}\right)^{-3}\left(\frac{m_\phi}{10^{-5}\textrm{ eV}}\right)^{-3}\textrm{ m}^3 \\
\textrm{Surface area:}&\hspace{1.5cm}A=4\pi^2\langle r\rangle\Delta r=120\left(\frac{\alpha}{0.037}\right)^{-2}\left(\frac{m_\phi}{10^{-5}\textrm{ eV}}\right)^{-2}\textrm{ m}^2.
\end{align}
\end{widetext}
In the $2p$-state, axions grow superradiantly at a rate
\be \Gamma_s=\frac{c^2\tilde{a}\alpha^8m_\phi}{24\hbar}=2.2*10^{-3}\tilde{a}\left(\frac{\alpha}{0.037}\right)^{8}\left(\frac{m_\phi}{10^{-5}\textrm{ eV}}\right)\textrm{ s}^{-1}, \ee
where $\tilde{a}$ is the black hole's spin parameter. At the same time, axions decay into photons at a rate
\be \Gamma_\phi=1.1*10^{-49}K^2\left(\frac{m_\phi}{10^{-5}\textrm{ eV}}\right)^{5}\textrm{ s}^{-1}, \ee
where $K$ is a model-dependent factor. For the more general axion-like particle, $K$ is a free parameter, but for the QCD axion, which we examine in this paper, $K\sim\mathcal{O}(1)$. Due to the quasi-bound nature of the $2p$-state, axions will not leave the $2p$-cloud except when disturbed by outside forces. However, a photon produced inside the $2p$-cloud will naturally exit the $2p$-cloud at the speed of light, producing an escape rate of
\be \Gamma_\gamma=\frac{c}{\Delta r}=2.5*10^8\left(\frac{\alpha}{0.037}\right)\left(\frac{m_\phi}{10^{-5}\textrm{ eV}}\right)\textrm{ s}^{-1}. \ee

\cite{Rosa_2018} found that the Boltzmann equations for the total number of photons and axions in the $2p$-cloud to be
\begin{align}
\frac{\d N_\phi}{\d t}&=\Gamma_sN_\phi-\Gamma_\phi\left(N_\phi(1+C_1N_\gamma)-C_2N_\gamma^2\right) \\
\frac{\d N_\gamma}{\d t}&=-\Gamma_\gamma N_\gamma+2\Gamma_\phi\left(N_\phi(1+C_1N_\gamma)-C_3N_\gamma^2\right),
\end{align}
where
\begin{align}
C_1&=\frac{8\alpha^2}{25} \hspace{1cm} C_2=\frac{2\alpha^4}{75} \hspace{1cm} C_3=C_2+\alpha C_1.
\end{align}
$\Gamma_\gamma\gg\Gamma_s\gg\Gamma_\phi$, and so for low values of $N_\gamma$ and $N_\phi$ photon production is negligible. In this regime, $N_\phi$ grows exponentially, $N_\phi\sim e^{\Gamma_st}$, while $N_\gamma\sim\frac{2\Gamma_\phi}{\Gamma_\gamma}N_\phi$. However, when $N_\gamma\gtrsim\frac{1}{C_1}$, the $C_1N_\gamma$ terms become significant, which increases the rate at which axions decay into photons; this marks the beginning of lasing. In this regime, $N_\gamma$ increases rapidly, while $N_\phi$'s growth slows, until both reach their equilibrium values:
\begin{align}
N^{\textrm{eq}}_\gamma&=\frac{\Gamma_s}{C_1\Gamma_\phi}=4.8*10^{49}\tilde{a}K^{-2}\left(\frac{\alpha}{0.037}\right)^6\left(\frac{m_\phi}{10^{-5}\textrm{ eV}}\right)^{-4} \\
N^{\textrm{eq}}_\phi&=\frac{\Gamma_\gamma}{2C_1\Gamma_\phi}=2.7*10^{60}K^{-2}\left(\frac{\alpha}{0.037}\right)^{-1}\left(\frac{m_\phi}{10^{-5}\textrm{ eV}}\right)^{-4}.
\end{align}

One important consideration is that a BLAST is only possible if the black hole has enough spin to produce the requisite number of axions. Since the $2p$-state has magnetic quantum number $m=1$, each axion produced by superradiance carries with it angular momentum $\hbar$, and therefore at the onset of lasing the axion cloud has angular momentum $\frac{\hbar\Gamma_\gamma}{2C_1\Gamma_\phi}$, whereas the black hole's angular momentum is given by $\tilde{a}\alpha\frac{M_{BH}}{m_\phi}$. This yields the requirement
\be \frac{0.06}{\tilde{a}\alpha^3K^2}\left(\frac{m_\phi}{10^{-8}\textrm{ eV}}\right)^{-2}\lesssim1. \ee
Superradiance requires $\alpha\lesssim.5$, and therefore we must have $m_\phi\gtrsim10^{-8}\textrm{ eV}$ and $M_{BH}\lesssim10^{-2}M_\odot$. This constraint restricts BLASTs to occur only around primordial Kerr black holes. Primordial black holes are expected to form with some small initial spin \cite{Chiba_2017,Harada_2017,Luca_2019,He_2019,Mirbabayi_2020,Harada_2021}, and they may spin up via a number of processes, including mergers \cite{Hofmann_2016}, accretion \cite{Ricotti_2007,Luca_2020}, hyperbolic encounters \cite{Jaraba_2021}, and Hawking radiation \cite{Taylor:2024fvf}. Thus it is possible that the conditions for a BLAST might form in nature.

In the process of reaching the equilibrium values listed earlier, the photon number reaches a maximum value of\begin{widetext}
\be N_\gamma^{\textrm{max}}\approx\frac{\Gamma_s}{C_1\Gamma_\phi}\ln\frac{\Gamma_s}{\Gamma_\phi}=5.1*10^{51}\tilde{a}K^{-2}\left(\frac{\alpha}{0.037}\right)^6\left(\frac{m_\phi}{10^{-5}\textrm{ eV}}\right)^{-4}\left(\frac{\xi}{106}\right), \ee
where
\be \xi=\ln\frac{\Gamma_s}{\Gamma_\phi}=107-4\ln\left(\frac{m_\phi}{10^{-5}\textrm{ eV}}\right)+8\ln\left(\frac{\alpha}{0.037}\right)+\ln\left(\frac{\tilde{a}}{K^2}\right). \ee
This corresponds to a peak luminosity of
\be L^{\textrm{max}}=\frac{m_\phi}{2}N_\gamma^{\textrm{max}}\Gamma_\gamma=1.0*10^{43}\tilde{a}K^{-2}\left(\frac{\alpha}{0.037}\right)^7\left(\frac{m_\phi}{10^{-5}\textrm{ eV}}\right)^{-2}\left(\frac{\xi}{106}\right)\textrm{ erg/s}. \ee\end{widetext}
This luminosity is comparable to that of the entire Milky Way galaxy. These BLASTs, if they exist, are therefore of great experimental interest, as they should be observable and could provide insight into both primordial black holes and scalar dark matter.

\section{electron-positron Plasmas in BLASTs}

\cite{Rosa_2018,Bo_kovi__2019} showed that the number of photons generated by a BLAST is so great that the electric field inside the $2p$-cloud will approach the critical Schwinger field. It is expected that, in this limit, the production of electron-positron pairs via the Schwinger effect will be significant. In the resulting electron-positron plasma, the photon has an effective mass equal to the plasma frequency. For a sufficient density of electrons and positrons, this will result in axions being energetically incapable of decaying into photons. At the same time, the dynamics of the plasma are nontrivial. In this section, we explore the behavior of this plasma, and the effect it has on axion decay. From this point onward, we will work in natural units where $c=\hbar=\epsilon_0=1$.

\subsection{Rate of Schwinger production}

The Schwinger effect \cite{PhysRev.82.664,Gelis_2016} is a nonperturbative effect caused by vacuum decay in the presence of an electromagnetic field. The Schwinger effect cannot occur when $E^2-B^2=\vec{E}\cdot\vec{B}=0$, and therefore a single beam of light cannot induce the Schwinger effect. However, when multiple beams of light intersect, the interference of the two electromagnetic fields may satisfy the requirements for Schwinger production \cite{Ringwald_2001,Alkofer_2001}; this is an area of significant research interest, as it is hoped that the Schwinger effect may be observable at the focus of two lasers (for a review, see \cite{10.1016/j.mre.2017.07.002}). In the same way, Schwinger production may occur inside the $2p$-cloud, where a large number of photons are traveling in different directions.

The Schwinger effect is driven not only by the electromagnetic field, but by the axion field as well \cite{Domcke_2021,Domcke_2022}. This is a consequence of the coupling between axions and electrons, which takes the form
\be \Lagr\supset\frac{g_{ae}}{2m_e}\bar{\psi}_e\gamma^\mu\gamma^5\psi_e\partial_\mu\phi, \ee
where $g_{ae}$ is the axion-electron coupling constant, $m_e$ is the electron mass, $\phi$ is the axion field, and $\psi_e$ is the electron bispinor. The presence of axions has the effect of enhancing the rate of pair production, compared to the Schwinger effect without axions; this is referred to as the axion-assisted Schwinger effect \cite{Domcke_2021,Domcke_2022}. This is an important consideration, however, for reasons that will be given momentarily, we do not include it in this paper, instead reserving it for future research. Therefore, the rate of Schwinger production given herein must be understood as a lower bound, as we are neglecting an effect that is known to enhance pair production. Put another way, the rate of Schwinger production that is used in this paper may be understood as the correct value only when $g_{ae}=0$; for $g_{ae}>0$, the rate of Schwinger production will be greater than what is used in this paper. Future research will be needed to determine the rate of Schwinger production when $g_{ae}>0$.

In the following subsection, we discuss the rate of Schwinger production assuming no axion assistance; after that we will discuss the difficulties presented by axion assistance. Finally, we present an alternative way in which the rate of pair production may be partially derived, in agreement with our result from the first subsection.

\subsubsection{The Schwinger effect in Compton-volume patches}

The Schwinger effect is typically calculated in the frame of reference in which the electric and magnetic field are parallel, which is guaranteed to exist except in the case where $E^2-B^2=\vec{E}\cdot\vec{B}=0$. The strength of the electric and magnetic fields in this frame of reference can be found by taking advantage of the Lorentz invariance of $E^2-B^2$ and $\vec{E}\cdot\vec{B}$:
\begin{align}
E_\parallel&=\sqrt{\frac{1}{2}\left(\sqrt{\left(E^2-B^2\right)^2+4(\vec{E}\cdot\vec{B})^2}+\left(E^2-B^2\right)\right)} \\
B_\parallel&=\sqrt{\frac{1}{2}\left(\sqrt{\left(E^2-B^2\right)^2+4(\vec{E}\cdot\vec{B})^2}-\left(E^2-B^2\right)\right)}.
\end{align}
For a constant, homogenous electromagnetic field, the rate of Schwinger production is
\be\label{ppconst} \frac{\d n_{e^+e^-}}{\d t}=\frac{e^2E_\parallel B_\parallel}{4\pi^2}\sum_{n=1}^\infty\frac{1}{n}\coth\frac{n\pi B_\parallel}{E_\parallel}e^{-n\pi\frac{E_c}{E_\parallel}}, \ee
where $E_c=\frac{m_e^2}{e}=1.3*10^{18}\textrm{ V/m}$ is the critical Schwinger field. Note that this increases with $E_\parallel$ but decreases with $B_\parallel$. In general, the Schwinger effect is highly dependent on the space- and time-dependence of the electromagnetic field, i.e. a nonconstant and/or nonuniform field will produce electrons and positrons at a very different rate than the figure given above. However, the effects of nonconstancy and nonuniformity are most pronounced when the electromagnetic field varies on length- and time-scales less than the Compton wavelength of an electron, $\lambda_C$ \cite{hebenstreit2011schwinger}.

In our case, the random decay of axions into photons results in a superposition of an enormous number of electromagnetic fields, so that the total electromagnetic field within the $2p$-cloud will be highly inhomogeneous and impossible to predict. However, the length- and time-scales of these inhomogeneities will be the wavelength of the light, which is $\frac{2}{m_\phi}$. For all axion masses that are of interest as a dark matter candidate, this is far greater than the Compton wavelength. Therefore we may divide the $2p$-cloud into a number of patches, each with volume equal to the Compton volume $\lambda_C^3$, and in each patch we may treat the electromagnetic field as uniform and constant. (It should be noted that most research on the Schwinger effect at the intersection of laser beams is done under the assumption of high-frequency light, typically X-rays, and in such cases this approximation does not hold. We are taking advantage of the fact that the light in a BLAST is at a much lower frequency than what would be practical in an earth-based experiment.) Therefore, the total rate of Schwinger pair production throughout the $2p$-cloud is
\be \frac{\d N_{e^+e^-}}{\d t}=\sum_{\textrm{patches}}\lambda_C^3\frac{e^2E_\parallel B_\parallel}{4\pi^2}\sum_{n=1}^\infty\frac{1}{n}\coth\frac{n\pi B_\parallel}{E_\parallel}e^{-n\pi\frac{E_c}{E_\parallel}}. \ee

If one wished, one could now find the average value of $E_\parallel$ and $B_\parallel$ in the $2p$-cloud, and substitute these values into the above equation. However, this approach would not produce an accurate estimate of the rate of Schwinger production. This is because of the Schwinger effect's highly nonlinear nature, meaning that even slight increases to $E_\parallel$ or decreases to $B_\parallel$ may result in a dramatic increase to the rate of pair production. This is significant in our case, because we have an electromagnetic field that is, effectively, randomly varying, and as such there will be some patches where the electromagnetic field produces electron-positron pairs at a greater or lesser rate than would be predicted by the cloud-wide averages of $E_\parallel$ and $B_\parallel$. These ``hotspots" and ``coldspots" contribute more than average and  less than average, respectively, to pair production. If the Schwinger effect were linear, the hotspots and coldspots would cancel out when averaging over the patches, and therefore they could be ignored. However, the Schwinger effect's nonlinear nature means that the surplus of pair production in the hotspots will be far greater than the deficit in pair production in the coldspots. Consequently, if we were to use the average values of $E_\parallel$ and $B_\parallel$, we would significantly underestimate the rate of pair production. What we do instead is find the probability distribution of the electric and magnetic field strength throughout the $2p$-cloud, $f(E)$ and $f(B)$, and average over these. Our total rate of pair production will then be
\begin{widetext}\be\label{Gammaschwdef} \frac{\d N_{e^+e^-}}{\d t}=V\int_0^\pi\frac{\d\theta\sin\theta}{2}\int_0^\infty\int_0^\infty\d E\d Bf(E)f(B)\frac{e^2E_\parallel B_\parallel}{4\pi^2}\sum_{n=1}^\infty\frac{1}{n}\coth\frac{n\pi B_\parallel}{E_\parallel}e^{-n\pi\frac{E_c}{E_\parallel}}\equiv\Gamma_{\textrm{Schw}}, \ee\end{widetext}
where $\theta$ is the angle between the electric and magnetic fields (assumed to be uniformly distributed between $0$ and $\pi$), and $E_\parallel$ and $B_\parallel$ are functions of $E$, $B$, and $\theta$. Note that $E$ and $B$ are the electric and magnetic field strengths in the axion cloud's zero-momentum frame, whereas $E_\parallel$ and $B_\parallel$ are the electric and magnetic field strengths in a different frame of reference (that being the one in which the electric and magnetic fields are parallel).

To find $f(E)$ and $f(B)$, we assume that the electromagnetic field components are normally distributed uncorrelated random variables with equal standard deviation, $\sigma_{\textrm{EM}}$. The electric field magnitude $E=\sqrt{E_x^2+E_y^2+E_z^2}$ then follows a chi distribution, with probability density
\be f(E)=\frac{\sqrt{2}}{\sqrt{\pi}\sigma_{\textrm{EM}}^3}E^2e^{-\frac{E^2}{2\sigma_{\textrm{EM}}^2}}. \ee
An analogous expression for $f(B)$ may be found in the same way. $\sigma_{\textrm{EM}}$ may be found from the energy density, which in terms of the electromagnetic field components is
\be u=\frac{1}{2}\left(E_x^2+E_y^2+E_z^2+B_x^2+B_y^2+B_z^2\right), \ee
and therefore
\be\label{sigmafind} \langle u\rangle=3\sigma_{\textrm{EM}}^2=\frac{m_\phi N_\gamma}{2V}. \ee
We perform the integral in eq.~\ref{Gammaschwdef} in Appendix \ref{gammacalc} and show that
\be\label{Schwrate} \Gamma_{\textrm{Schw}}\approx\frac{e^2E_c^2V}{8\sqrt{3}\pi}\sum_{n=1}^\infty e^{-3\left(\frac{N_\gamma}{n^2N_\gamma^{\textrm{Schw}}}\right)^{\frac{-1}{3}}}\left(\frac{N_\gamma}{n^2N_\gamma^{\textrm{Schw}}}\right)^{\frac{1}{3}} \ee
where
\be\label{ngammaschw} N_\gamma^{\textrm{Schw}}=\frac{3\pi^2E_c^2V}{4m_\phi}=4.9*10^{51}\left(\frac{\alpha}{0.037}\right)^{-3}\left(\frac{m_\phi}{10^{-5}\textrm{ eV}}\right)^{-4}. \ee
This approximation holds for $N_\gamma\lesssim.1N_\gamma^{\textrm{Schw}}$. We will find that, for our purposes, this is always the case, so that we may freely use this approximation for $\Gamma_{\textrm{Schw}}$.

\subsubsection{The challenges of axion-assistance}

In principle, the approach we've taken can easily be extended to account for axion-assistance. The axion field will vary in three ways with respect to time and space: it has a time-dependent phase factor $e^{-im_\phi t}$, its magnitude increases with time as a consequence of superradiant growth, and it is spatially inhomogeneous, which results from solving the Klein-Gordon equation in Kerr spacetime. These spatial inhomogeneities have lengthscale $\frac{2}{\alpha m_\phi}$ \cite{Detweiler:1980uk}, and superradiant growth occurs over timescale $\frac{1}{2\Gamma_s}$. The shortest timescale is that which corresponds to the phase factor, $\frac{1}{m_\phi}$, and therefore this is the dominant form of axion assistance (since the axion couples to the electron via its derivative, $\partial_\mu\phi$). All of these length- and timescales are far larger than $\lambda_C$, and so, when we divide the axion cloud into Compton volumes, we may consider only the dominant term and approximate
\be \phi=e^{-im_\phi t}\lvert\phi\rvert. \ee
In principle, all we need do is derive an analog to eq.~\ref{ppconst} for the case of $\phi=e^{-im_\phi t}\lvert\phi\rvert$, i.e. find the rate of Schwinger production when the electric field, magnetic field, and magnitude of the axion field are all held constant. We may then repeat the same procedure as before, inserting this analog of eq.~\ref{ppconst} into eq.~\ref{Gammaschwdef}.

The challenge here lies in finding this analog to eq.~\ref{ppconst}. To the author's knowledge, doing so analytically is not a solved problem in the literature; previous works on the axion-assisted Schwinger effect \cite{Domcke_2021,Domcke_2022} have been primarily numerical in nature. The prescription in these papers is to write a differential equation governing the time evolution of the Bogoliubov coefficients for the fermion, and then to numerically solve this differential equation. This yields the rate of pair production at a single point in momentum-space. To arrive at the total rate of pair production within a single Compton-volume patch (i.e., our analog to eq.~\ref{ppconst}), one would need to integrate over momentum space. Then, to find the total rate of pair production throughout the entire $2p$-cloud, one would need to integrate over $E$, $B$, and $\theta$, analogous to eq.~\ref{Gammaschwdef}. This would yield the rate of pair production in the $2p$-cloud, but only for a single value of $N_\gamma$ and $N_\phi$. Because $N_\gamma$ and $N_\phi$ will change over time, one would need to repeat this procedure as the axion laser evolves with time. The upshot is that numerically modeling the impact of the axion-assisted Schwinger effect on a superradiant axion laser would require an octuple integral (one integral to find the Bogoliubov coefficients, three over momentum space, three over the electric and magnetic fields, and one over time). As this is unlikely to be computationally feasible, it is necessary to determine an analytical expression to serve as our eq.~\ref{ppconst}-analog.

Deriving the rate of Schwinger production in the presence of both a constant electromagnetic field and an axion field that is constant up to a phase factor, is a sufficiently significant problem to deserve being a paper unto itself. For this paper, we defer that problem, neglecting axion assistance but noting that our expression for $\Gamma_{\textrm{Schw}}$ nevertheless has value as a lower bound. A more complete understanding of the Schwinger effect in BLASTs, applicable to the case of $g_{ae}>0$, will only be possible when the aforementioned eq.~\ref{ppconst}-analog has been derived.

\subsubsection{An alternate perspective on pair production}

In the preceding subsections, we approached this problem from the perspective of the classical electromagnetic field; we argued that, for sufficiently large $N_\gamma$, the field is sufficient to induce pair production at a significant rate. There is an alternative way of considering the problem, which is to focus on the individual photons present in the $2p$-cloud, produced by axion decay. This is an attractive lens through which to view the situation, as we know to a good approximation the spectrum and distribution of these photons (i.e., monochromatic, and uniformly distributed both in space and propagation direction). At the level of the individual photons, the Schwinger effect is caused by interactions between photons and a single electron-positron loop. It is an interesting question to see if we can derive our earlier results from the perspective of these specific particle interactions.

\cite{Martin_2003} calculates the rate at which $N$ photons, $K$ of which have positive helicity, will undergo a scattering interaction involving a single loop. They find that, within an electromagnetic field defined by the tensor $F_{\mu\nu}$, such a scattering interaction will take place at the rate
\be \Gamma_{N,K}=-\frac{m_e^4}{8\pi^2}\left(\frac{2e}{m_e^2}\right)^Nc\left(\frac{K}{2},\frac{N-K}{2}\right)\chi_+^{\frac{K}{2}}\chi_-^{\frac{N-K}{2}}, \ee
where $c$ is a function defined in \cite{Martin_2003}, and $\chi_\pm$ are related to the electromagnetic field:
\be \chi_\pm=\frac{1}{8}F_{\mu\nu}F^{\mu\nu}\pm\frac{i}{8}F_{\mu\nu}\tilde{F}^{\mu\nu}. \ee
Without knowing the specific form of $F_{\mu\nu}$ (which will vary as a function of time and position), by considering the energy density of the electromagnetic field we may say that, for a cloud of photons with angular frequency $\omega$ and number density $n_\gamma$, we will have
\be F_{\mu\nu}=\sqrt{n_\gamma\omega}f_{\mu\nu}(\vec{x},t), \ee
where $f_{\mu\nu}$ is a dimensionless function of time and space. The rate of scattering will therefore be
\be \Gamma_{N,K}\propto m_e^4\left(\frac{e^2\omega}{m_e^4}n_\gamma\right)^{\frac{N}{2}}, \ee
where the constant of proportionality will depend on $N$ and $K$, as well as time and position within the photon cloud. If $\frac{e^2\omega}{m_e^4}n_\gamma\ll1$, $\Gamma_{N,K}$ will vanish for large $N$, and the rate of one-loop interaction will be negligible. Conversely, there exists some critical number density of photons above which pair production is significant, given by
\be n_\gamma^{\textrm{crit}}\sim\frac{m_e^4}{e^2\omega}. \ee
Bearing in mind that $E_c=\frac{m_e^2}{e}$ and $\omega=\frac{m_\phi}{2}$, this reproduces, up to a constant of proportionality, eq.~\ref{ngammaschw}.

\subsection{Positron/Electron Escape Rate}

Just as photons will exit the $2p$-cloud at a rate $\Gamma_\gamma$, we also must calculate the rate at which electrons and positrons will exit the $2p$-cloud. Schwinger pairs are created with some initial momentum, but we assume that this is negligible compared to the momentum imparted by radiation pressure. This radiation pressure is driven by an energy flux, which we assume to be constant throughout the $2p$-cloud and which is given by $F=\frac{L}{A}=\frac{3m_\phi}{4V}N_\gamma$. Let $\rho$ be a coordinate defined as the distance from the ring running through the center of the $2p$-torus; we expect that radiation pressure will cause electrons and positrons to move predominantly in the $\hat{\rho}$ direction. Therefore each electron and positron gains energy at a rate  
\begin{widetext}\be \frac{3\sigma_Tm_\phi}{4V}N_\gamma=\frac{\d}{\d t}g_{tt}p^t=\frac{\d}{\d t}\left(\left(1-\frac{r_sr}{r^2+\frac{1}{4}r_s^2\tilde{a}^2\cos^2\theta}\right)\frac{m_e}{\sqrt{1-\left(\frac{\d\rho}{\d t}\right)^2}}\right), \ee
where $g$ is the Kerr metric, $p$ is the four-momentum (related to the energy by $E=p_t$), $r_s$ is the Schwarzschild radius of the black hole, and $r$ and $\theta$ are Boyer-Lindquist coordinates. The factor of $g_{tt}$ accounts for the curvature of Kerr spacetime; note that there could also be a $g_{t\phi}p^\phi$ term, corresponding  to the electron or positron's azimuthal motion, except for the fact that, if the electron or positron is moving primarily in the $\hat{\rho}$ direction, then $p^\phi$ is negligible. The Schwarzschild radius may be written as
\be r_s=\frac{2GM}{c^2}=1.5*10^{-3}\left(\frac{\alpha}{.037}\right)\left(\frac{m_\phi}{10^{-5}\textrm{ eV}}\right)^{-1}\textrm{ m}, \ee
and therefore for all superradiant modes ($\alpha\lesssim.5$), $r\gg r_s$ at all points within the $2p$-cloud. We may therefore neglect this prefactor; put another way, the $2p$-cloud is far enough away from the black hole that we may treat spacetime as flat.

We assume that the timescale on which an individual electron or positron accelerates is much shorter than the timescale over which $N_\gamma$ changes, and therefore we may treat $N_\gamma$ as constant. From this we find that the time $T_{e^\pm}$ that it takes for the electron or positron to leave the $2p$-cloud is given by
\begin{align}
\Delta r-\rho_0&=\sqrt{T_{e^\pm}\left(\frac{8Vm_e}{3\sigma_Tm_\phi N_\gamma}+T_{e^\pm}\right)} \nonumber\\
&\hspace{1cm}+\frac{4Vm_e}{3\sigma_Tm_\phi N_\gamma}\left(\frac{\pi}{3}-2\tan^{-1}\frac{\sqrt{3}\left(\frac{4Vm_e}{3\sigma_Tm_\phi N_\gamma}+T_{e^\pm}\right)+2\sqrt{T_{e^\pm}\left(\frac{8Vm_e}{3\sigma_Tm_\phi N_\gamma}+T_{e^\pm}\right)}}{\frac{4Vm_e}{\sigma_Tm_\phi N_\gamma}+T_{e^\pm}}\right).
\end{align}
\end{widetext}
where $\rho_0$ is the value of $\rho$ where the electron or positron was created. While it is not possible to get an exact analytical expression for $T_{e^\pm}$ in terms of $\rho_0$, the above equation can be numerically inverted, so that in principal we may calculate the escape rate as
\be \Gamma_{e^\pm}=\frac{1}{\langle T_{e^\pm}\rangle}=\frac{\Delta r^2}{2\int_0^{\Delta r}T_{e^\pm}\rho_0\d\rho_0}. \ee
Notably, in the limit $\frac{3\sigma_Tm_\phi}{4Vm_e}\Delta rN_\gamma\to\infty$, $\Gamma_{e^\pm}\to\Gamma_\gamma$; this represents the limit where the radiation pressure is so great that it immediately accelerates electrons and positrons to relativistic velocities. We may therefore use
\be \chi\equiv\frac{3\sigma_Tm_\phi}{4Vm_e}\Delta rN_\gamma=\frac{N_\gamma}{6.2*10^{40}}\left(\frac{\alpha}{0.037}\right)^{2}\left(\frac{m_\phi}{10^{-5}\textrm{ eV}}\right)^{3} \ee
\hfill\\
\noindent as a dimensionless parameter for the behavior of $\Gamma_{e^\pm}$. $\Gamma_{e^\pm}$ increases monotonically as $\chi$ increases, asymptotically approaching $\Gamma_\gamma$. This behavior is analyzed in greater detail in Appendix \ref{chianalysis}, and it is plotted in Fig. 1a.

\begin{figure*}
\includegraphics[scale=.7]{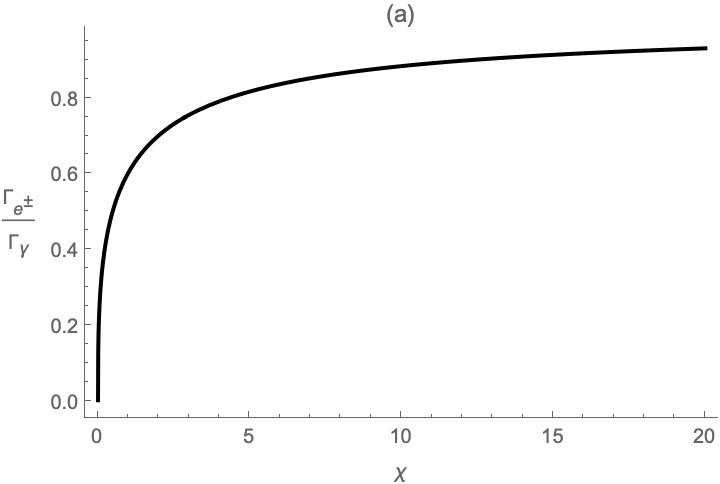} \\
\hspace{1cm}\\
\hspace{-.5cm}\includegraphics[scale=.75]{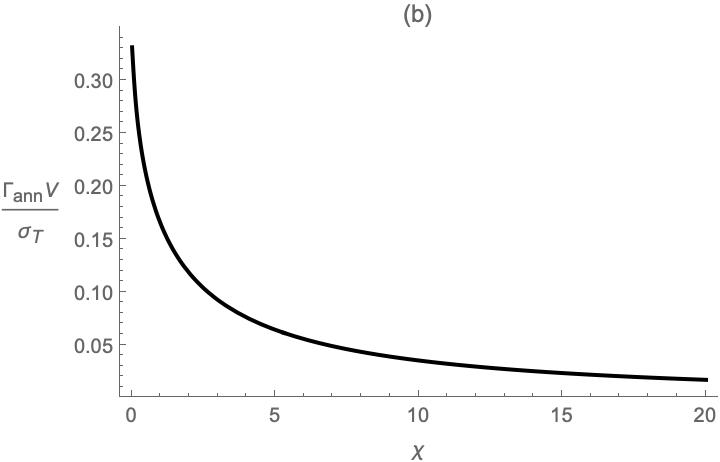} \\
\hspace{1cm}\\
\includegraphics[scale=.7]{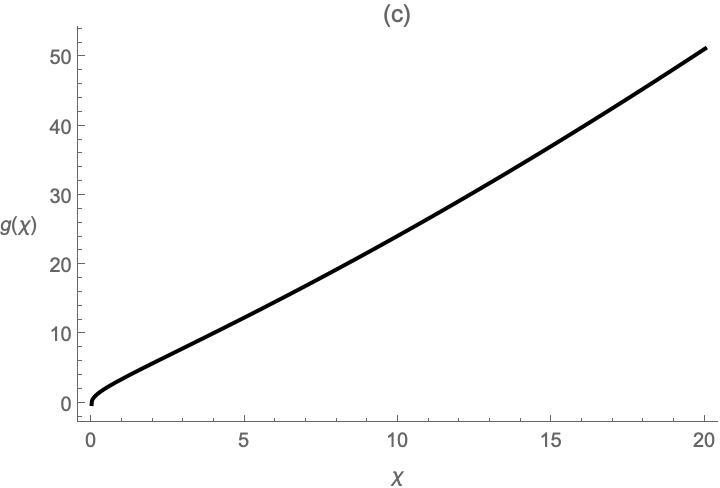}
\caption{Plot of functions of $\chi$, which parameterizes how relativistic the electrons and positrons become as a result of radiation pressure. (a) demonstrates the behavior of $\Gamma_{e^\pm}$. At low $\chi$, there is negligible radiation pressure driving electrons and positrons to exit the $2p$-cloud, and so $\Gamma_{e^\pm}\to0$. For large $\chi$ the electrons and positrons are quickly accelerated to very close to the speed of light, causing them to exit the $2p$-cloud at roughly the same rate as the photons, and as a result $\Gamma_{e^\pm}\to\Gamma_\gamma$. (b) demonstrates the behavior of $\Gamma_{\textrm{ann}}$. Slower particles present a larger cross-section, and therefore pair annihilation occurs frequently for small $\chi$ but vanishes as $\chi\to\infty$. Note that $\Gamma_{\textrm{ann}}$ is the rate of annihilation per number of pairs squared, not the total rate; in practice, large $\chi$ will correlate with large $N_{e^+e^-}$, so the total rate of annihilation will increase with $\chi$. (c) demonstrates the behavior of $g(\chi)$, which scales the number of electron-positron pairs at which annihilation becomes relevant. We show in Appendix \ref{chianalysis} that, for large $\chi$, $g(\chi)$ grows as $\frac{\chi^2}{(\ln\chi)^2}$.}
\end{figure*}

\subsection{Pair Annihilation}

Within the electron-positron plasma, some number of electrons and positrons will annihilate and produce two photons. The cross-section of this annihilation is given by \cite{1982ApJ...258..321S,1990MNRAS.245..453C}
\begin{widetext}\be \sigma(v_{\textrm{com}})=\frac{3\sigma_T(1-v_{\textrm{com}})^2}{32v_{\textrm{com}}}\left(\frac{3-v_{\textrm{com}}^4}{v_{\textrm{com}}}\ln\frac{1+v_{\textrm{com}}}{1-v_{\textrm{com}}}-2(2-v_{\textrm{com}}^2)\right), \ee\end{widetext}
where $v_{\textrm{com}}$ is the velocity of the electron and positron in the center-of-mass frame. Due to the effects of radiation pressure, described in the previous section, we expect the electrons and positrons are primarily moving in the $\hat{\rho}$ direction; we may therefore treat the electron-positron plasma as a series of beams, each pointed in the $\hat{\rho}$ direction. For a pure beam of positrons and a pure beam of electrons with velocities $v_{e^+}$ and $v_{e^-}$, respectively, each of which is pointed in the same direction, the rate of annihilation is
\be \frac{\d n_{e^\pm}}{\d t}=-n_{e^+}n_{e^-}\lvert v_{e^+}-v_{e^-}\rvert\sigma(v_{\textrm{com}}), \ee
The center-of-mass velocity is related to the electron's and positron's velocities by
\be v_{\textrm{com}}=\sqrt{\frac{\gamma_{e^+}\gamma_{e^-}(1-v_{e^+}v_{e^-})-1}{\gamma_{e^+}\gamma_{e^-}(1-v_{e^+}v_{e^-})+1}}, \ee
where $\gamma_{e^\pm}$ are the Lorentz factors corresponding to $v_{e^\pm}$.

In our case, the electrons and positrons have a distribution of velocities, $f(v)$, and therefore the total rate of annihilation is
\begin{align}
\frac{\d n_{e^\pm}}{\d t}=-n_{e^+}n_{e^-}\int dv_{e^+}&dv_{e^-} f(v_{e^+})f(v_{e^-}) \nonumber\\
&\times\lvert v_{e^+}-v_{e^-}\rvert\sigma(v_{\textrm{com}}).
\end{align}
Using the same approach as in the previous section, the velocity of an electron or positron is directly related to the length of time it has existed, $t$:
\be t=\frac{1}{\sqrt{1-v^2}}\frac{4m_eV}{3\sigma_Tm_\phi N_\gamma}=\frac{\Delta r}{\chi\sqrt{1-v^2}}, \ee
and therefore we may relate the velocity of the electron or positron to the distance it has traveled since it was created, which we call $\Delta\rho$:
\be \Delta\rho=\frac{\Delta r}{\chi}\left(\frac{v}{\sqrt{1-v^2}}-\tan^{-1}\frac{v}{\sqrt{1-v^2}}\right). \ee
As in the previous section, we cannot write an analytic expression for $v$ in terms of $\Delta\rho$, but it is still possible in principle to express the former as a function of the latter, $v(\Delta\rho)$. We may therefore replace our integral over the velocities of the electron and positron with an integral over the locations at which the two particles are created and at which the annihilation occurs. Assuming that electrons and positrons are created uniformly throughout the $2p$-torus, and that annihilation occurs uniformly throughout the $2p$-torus, we get that the rate of annihilation is
\begin{widetext}\be \frac{\d n_{e^\pm}}{\d t}=-n_{e^+}n_{e^-}\int_0^{\Delta r}\d\rho\frac{2\rho}{\Delta r^2}\int_0^\rho\d\rho_{0e^+}\frac{2\rho_{0e^+}}{\rho^2}\int_0^\rho\d\rho_{0e^-}\frac{2\rho_{0e^-}}{\rho^2}\lvert v_{e^+}-v_{e^-}\rvert\sigma(v_{\textrm{com}}), \ee
where $v_{e^\pm}$ is a function of $\rho-\rho_{0e^\pm}$. In this integral, $\rho$ represents the coordinate at which annihilation occurs, and $\rho_{0e^\pm}$ represents the coordinate at which the electron or positron was created. The factors of $\frac{2\rho}{\Delta r^2}$ and $\frac{2\rho_{0e^\pm}}{\rho^2}$ are the probability distributions for $\rho$ and $\rho_{0e^\pm}$. Since the number of electrons and positrons is assumed to be equal, we may write this as
\be \frac{\d N_{e^+e^-}}{\d t}=-\frac{8N_{e^+e^-}^2}{V\Delta r^2}\int_0^{\Delta r}\d\rho\int_0^\rho\int_0^\rho\d\rho_{0e^+}\d\rho_{0e^-}\frac{\rho_{0e^+}\rho_{0e^-}}{\rho^3}\lvert v_{e^+}-v_{e^-}\rvert\sigma(v_{\textrm{com}})\equiv-\Gamma_{\textrm{ann}}N_{e^+e^-}^2. \ee\end{widetext}
We show in Appendix \ref{chianalysis} that $\Gamma_{\textrm{ann}}V$ is a function only of $\chi$, which is displayed in Fig. 1b. As $\chi$ increases, the electrons and positrons become more relativistic, and so both the cross section and the rate of annihilation decrease.

Each pair annihilation creates two photons, and therefore there ought to be a $\Gamma_{\textrm{ann}}N_{e^+e^-}^2$ term in the photon Boltzmann equation. However, these photons will have energy of at least $m_e$, whereas the photons created by axion decay have energy $\frac{m_\phi}{2}\ll m_e$. A rigorous analysis will therefore require us to keep track of two separate photon populations, low-energy photons produced by axion decay and high-energy photons produced by pair annihilation; the $\Gamma_{\textrm{ann}}N_{e^+e^-}^2$ term would appear in the Boltzmann equation of the latter, but not the former. However, as we will show later, pair annihilation does not occur at a significant rate, and therefore we may ignore these high-energy photons.

\subsection{Suppression of Axion Decay}

Lastly, we must account for the effects of the electron-positron plasma on axion decay. The plasma frequency is given by
\be \omega_p=\sqrt{\frac{2e^2N_{e^+e^-}}{m_eV}}, \ee
and this serves as the effective mass of photons in the $2p$-cloud. It is a straightforward exercise of QED to show that this scales the rate of axion decay by
\be \Gamma_\phi\to\Gamma_\phi\textrm{Re}\sqrt{1-\frac{8e^2}{m_em_\phi^2V}N_{e^+e^-}}\equiv\Gamma'_\phi. \ee
Note that this has the effect of shutting down axion decay entirely when
\begin{align}
N_{e^+e^-}&\geq\frac{m_em_\phi^2V}{8e^2}=6.5*10^{17}\left(\frac{\alpha}{0.037}\right)^{-3}\left(\frac{m_\phi}{10^{-5}\textrm{ eV}}\right)^{-1}\nonumber\\&\equiv N_{e^+e^-}^{\textrm{shutoff}}. \end{align}

Another consequence of the photon's effective mass is that it permits the photon to grow superradiantly \cite{Pani_2013,Conlon_2018,Cannizzaro_2021a,Cannizzaro_2021b,cannizzaro2023nonlinearphotonplasmainteractionblack}. This effect is not considered in this paper, as accounting for it would require an understanding of the spatial distribution of the plasma, which is beyond our present scope. There is reason to believe it may not be significant, as numerical simulations have shown that this superradiant growth becomes significant only when $\omega_p>\frac{m_\phi}{2}$ \cite{spieksma2023superradianceaxioniccouplingsplasma}, i.e. $N_{e^+e^-}>N_{e^+e^-}^{\textrm{shutoff}}$. We will show later that, to a good approximation, $N_{e^+e^-}$ increases monotonically with $N_\gamma$; therefore, because $N_\gamma$ will not increase when $N_{e^+e^-}\geq N_{e^+e^-}^{\textrm{shutoff}}$, we do not expect $N_{e^+e^-}$ to exceed $N_{e^+e^-}^{\textrm{shutoff}}$. It must be acknowledged, however, that this is only an initial prediction, and a more detailed numerical model may reveal the superradiant growth of the photon to be significant.

\section{Boltzmann Equations and Parameter Space}

We may now write the Boltzmann equations for the axion number, photon number, and Schwinger pair number:
\begin{align}
\frac{\d N_\phi}{\d t}&=\Gamma_sN_\phi-\Gamma'_\phi\left(N_\phi(1+C_1N_\gamma)-C_2N_\gamma^2\right) \\
\frac{\d N_\gamma}{\d t}&=-\Gamma_\gamma N_\gamma+2\Gamma'_\phi\left(N_\phi(1+C_1N_\gamma)-C_3N_\gamma^2\right) \nonumber\\
&\hspace{.6cm}-\frac{4m_e}{m_\phi}\Gamma_{\textrm{Schw}} \\
\frac{\d N_{e^+e^-}}{\d t}&=\Gamma_{\textrm{Schw}}-\Gamma_{e^{\pm}}N_{e^+e^-}-\Gamma_{\textrm{ann}}N_{e^+e^-}^2.
\end{align}
Note that, because we are modeling the total number of particles in the axion cloud, we have only included phenomena that can introduce or remove particles from the cloud. There are likely to be additional phenomena that alter the distribution of particles within the axion cloud (e.g. scattering between electrons and positrons), but these are not relevant to this paper.

Numerical solutions to these equations, for two different sets of parameters, are shown in Fig. 2. Initially, the rate of pair production is vanishingly small, so $N_\phi$ and $N_\gamma$ behave identically to the description given in \cite{Rosa_2018}. When $N_\gamma$ gets within a few orders of magnitude of $N_\gamma^{\textrm{Schw}}$ (roughly $10^{-4}N_\gamma^{\textrm{Schw}}$), electron-positron pairs begin to be produced, and these electron-positron pairs impact the axion decay rate. Thus we need only consider the case when $N_\gamma$ is approaching $N_\gamma^{\textrm{Schw}}$.

\begin{figure*}
\begin{tabular}{c}\hspace{1cm}(ai)\hspace{8cm}(aii)\\
\includegraphics[width=\textwidth]{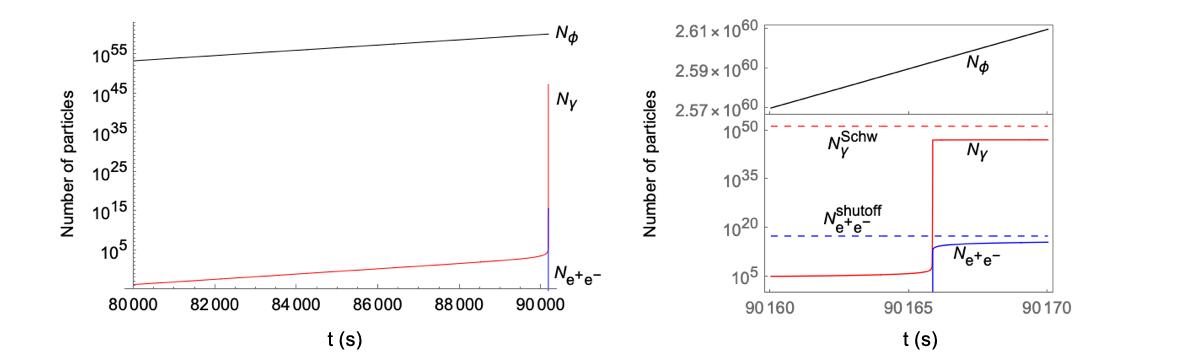}\\
\hspace{1cm}(bi)\hspace{8cm}(bii)\\
\includegraphics[width=\textwidth]{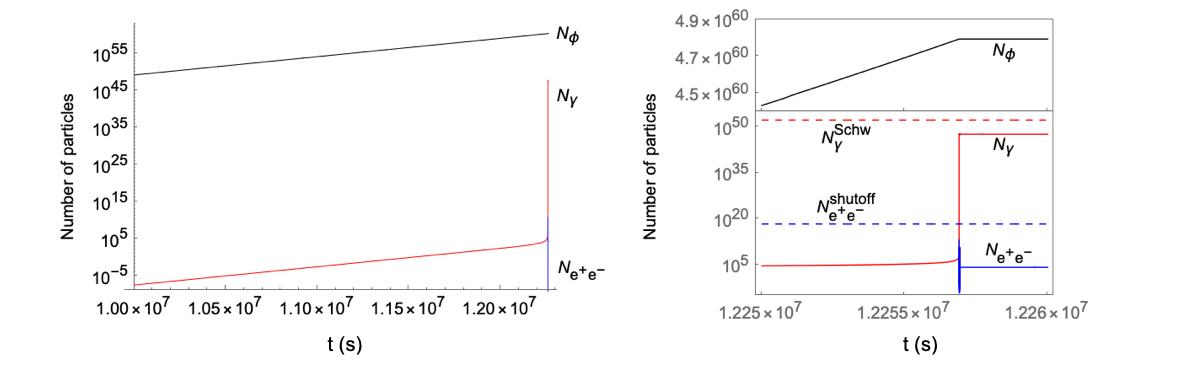}\end{tabular}
\caption{Numerical simulations of the growth of $N_\phi$, $N_\gamma$, and $N_{e^+e^-}$. (a) is done with $\alpha=.037$ while (b) is done with $\alpha=.02$, with the other parameters set to $m_\phi=10^{-5}\textrm{ eV}$, $\tilde{a}=.7$, and $K=1$. (aii) and (bii) show zoomed-in views of the onset of lasing, when $N_\gamma$ increases rapidly. In both cases, $N_{e^+e^-}$ increases alongside $N_\gamma$, matching with the approximation $N_{e^+e^-}\approx\frac{\Gamma_{\textrm{Schw}}}{\Gamma_{e^\pm}}$. Prior to the onset of lasing, $N_\phi$ and $N_\gamma$ increase exponentially, while $N_{e^+e^-}$ is negligible. In (a), $N_{e^+e^-}$ approaches $N_{e^+e^-}^{\textrm{shutoff}}$, slowing axion decay, so that $N_\gamma$ stops increasing while $N_\phi$ continues increasing. In (b), $N_{e^+e^-}$ is not large enough to significantly affect axion decay, and so $N_\gamma$ and $N_\phi$ settle into their equilibrium values. (a) lies in the unstable region of parameter space, while (b) lies in the unenhanced region (see Figs. 3 and 4). Note that, in both plots, $N_\gamma$ always remains multiple orders of magnitude below $N_\gamma^{\textrm{Schw}}$, allowing for the approximation given in eq.~\ref{Schwrate}.}
\end{figure*}

Let us first consider the behavior of $N_{e^+e^-}$. When $N_{e^+e^-}$ is small, the annihilation term is negligible, and so the number of electron-positron pairs is controlled only by the rate of production and the rate of escape. For all $N_\gamma<N_\gamma^{\textrm{Schw}}$, $\Gamma_{e^\pm}\gg\frac{\Gamma_{\textrm{Schw}}}{N_\gamma}$, and therefore we may approximate $N_{e^+e^-}\approx\frac{\Gamma_{\textrm{Schw}}}{\Gamma_{e^\pm}}$. This approximation holds until either annihilation becomes significant, or until $N_\gamma$ exceeds $N_\gamma^{\textrm{Schw}}$. We will show that, in what appears to be a remarkable coincidence, both of these conditions are met at roughly the same time. Examining first the matter of $N_\gamma$ exceeding $N_\gamma^{\textrm{Schw}}$, we note that, when $N_\gamma=N_\gamma^{\textrm{Schw}}$, $\chi$ is given by
\be \chi^{\textrm{Schw}}=7.8*10^{10}\left(\frac{\alpha}{0.037}\right)^{-1}\left(\frac{m_\phi}{10^{-5}\textrm{ eV}}\right)^{-1}, \ee
and the number of electron-positron pairs is, provided the aforementioned approximation still holds at this point,
\be N_{e^+e^-}^{\textrm{Schw}}=4*10^{48}\left(\frac{\alpha}{0.037}\right)^{-4}\left(\frac{m_\phi}{10^{-5}\textrm{ eV}}\right)^{-4}. \ee
As for pair annihilation, that becomes significant when $N_{e^+e^-}\sim\frac{\Gamma_{e^\pm}}{\Gamma_{\textrm{ann}}}\equiv N_{e^+e^-}^{\textrm{ann}}$, which may be written as\begin{widetext}
\be N_{e^+e^-}^{\textrm{ann}}=\frac{\Gamma_\gamma V}{\sigma_T}g(\chi)=2.7*10^{30}\left(\frac{\alpha}{0.037}\right)^{-2}\left(\frac{m_\phi}{10^{-5}\textrm{ eV}}\right)^{-2}g(\chi), \ee
where
\be g(\chi)=\frac{\chi^5}{48\left(\int_0^\chi \hat{T}_{e^\pm}\hat{\rho}_0\d\hat{\rho}_0\right)\left(\int_0^\chi\d\hat{\rho}\int_0^{\hat{\rho}}\d\hat{\rho}_{0e^+}\int_0^{\hat{\rho}}\d\hat{\rho}_{0e^-}\frac{\hat{\rho}_{0e^+}\hat{\rho}_{0e^-}}{\hat{\rho}^3}\lvert v_{e^+}-v_{e^-}\rvert\frac{\sigma(v_{\textrm{com}})}{\sigma_T}\right)}, \ee\end{widetext}
which uses the hat notation defined in Appendix \ref{chianalysis}. $g(\chi)$ is plotted in Fig. 1c, and we show in Appendix \ref{chianalysis} that its asymptotic form is
\be g(\chi)\xrightarrow[\chi\to\infty]{}\frac{2\chi^2}{3(\ln\chi)^2}. \label{gasymptote}\ee
This means that, at $\chi=\chi^{\textrm{Schw}}$, $N_{e^+e^-}^{\textrm{ann}}\sim N_{e^+e^-}^{\textrm{Schw}}$. In other words, pair annihilation only becomes relevant right as $N_\gamma$ reaches $N_\gamma^{\textrm{Schw}}$. Thus, in the $N_\gamma<N_\gamma^{\textrm{Schw}}$ regime, we may ignore annihilation, and the approximation $N_{e^+e^-}\approx\frac{\Gamma_{\textrm{Schw}}}{\Gamma_{e^\pm}}$ holds.

We now examine the assumption that $N_\gamma<N_\gamma^{\textrm{Schw}}$. The only positive term in the Boltzmann equation vanishes when $N_{e^+e^-}\geq N_{e^+e^-}^{\textrm{shutoff}}$. If this happens when $N_\gamma<N_\gamma^{\textrm{Schw}}$, then the photon number will have a ceiling, which we refer to as $N_\gamma^{\textrm{ceiling}}$ and which may be found by solving $\frac{\Gamma_{\textrm{Schw}}}{\Gamma_{e^\pm}}=N_{e^+e^-}^{\textrm{shutoff}}$ for $N_\gamma$. Note that $N_\gamma$ might never reach $N_\gamma^{\textrm{ceiling}}$. This inversion is easy to do numerically, but it is impossible analytically, and therefore it is unclear how $N_\gamma^{\textrm{ceiling}}$ depends on $\alpha$ and $m_\phi$. However, it is straightforward to show that, for any reasonable values of $\alpha$ and $m_\phi$, $\frac{\Gamma_{\textrm{Schw}}}{\Gamma_{e^\pm}}\gg N_{e^+e^-}^{\textrm{shutoff}}$ when $N_\gamma=N_\gamma^{\textrm{Schw}}$. From this we conclude that $N_\gamma^{\textrm{ceiling}}\ll N_\gamma^{\textrm{Schw}}$, and we may therefore assume, throughout this analysis, that $N_\gamma<N_\gamma^{\textrm{Schw}}$, and we may use the approximations derived from this assumption. 

Let us next turn to the effects of $N_{e^+e^-}$ on $N_\phi$ and $N_\gamma$. There are two conditions under which the electron-positron plasma may become significant to the dynamics of the photon number and the axion number. The first way is that, as $N_{e^+e^-}$ approaches $N_{e^+e^-}^{\textrm{shutoff}}$, the rate of axion decay $\Gamma'_\phi$ vanishes. This happens when $\frac{\Gamma_{\textrm{Schw}}}{\Gamma_{e^\pm}N_{e^+e^-}^{\textrm{shutoff}}}\sim1$. The second way is that, for sufficiently large $N_\gamma$, the rate of photon loss from the Schwinger effect will become significant. This happens when $\frac{4m_e\Gamma_{\textrm{Schw}}}{m_\phi\Gamma_\gamma N_\gamma}\sim1$. For all values of $N_\gamma$, $\frac{\Gamma_{\textrm{Schw}}}{\Gamma_{e^\pm}N_{e^+e^-}^{\textrm{shutoff}}}\gg\frac{4m_e\Gamma_{\textrm{Schw}}}{m_\phi\Gamma_\gamma N_\gamma}$, and therefore the former condition will be met before the latter. Once the number of electron-positron pairs reaches $N_{e^+e^-}^{\textrm{shutoff}}$, $N_\gamma$ ceases to increase, and consequently it will never become large enough for the second condition to be fulfilled. We may therefore neglect the $\frac{4m_e}{m_\phi}\Gamma_{\textrm{Schw}}$ term in the $N_\gamma$ Boltzmann equation.

\begin{figure}
\includegraphics[width=\columnwidth]{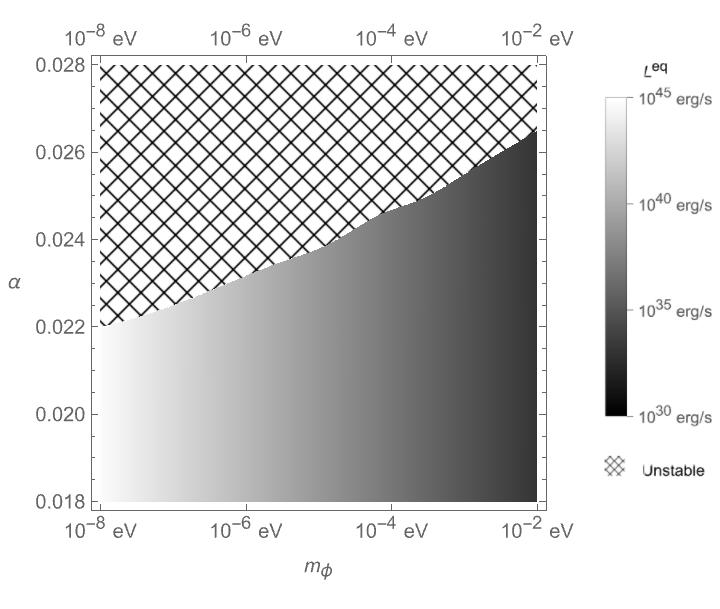}
\caption{Plot of the $\alpha m_\phi$ plane of parameter space. The other parameters were taken to be $\tilde{a}=.7$ and $K=1$. For larger values of $\alpha$ and smaller values of $m_\phi$, the system is unstable, represented by the black-colored region of parameter space. In this region, the electron-positron plasma suppresses the axion decay so much that the photon number never becomes large enough to halt the growth of the axion number. The axion number therefore continues to grow superradiantly, causing the black hole to spin down until it is no longer superradiant. In the colored regions of parameter space, it is possible for the system to reach equilibrium, and the equilibrium luminosity is plotted. Especially for light axions, these luminosities are extremely high, making BLASTs potentially significant for future experiments. In the shaded region of parameter space, $L^{\textrm{eq}}$ depends on both $\alpha$ and $m_\phi$, but the dependence on $\alpha$ is too small to see on this plot.}
\end{figure}

\begin{figure}
\includegraphics[width=\columnwidth]{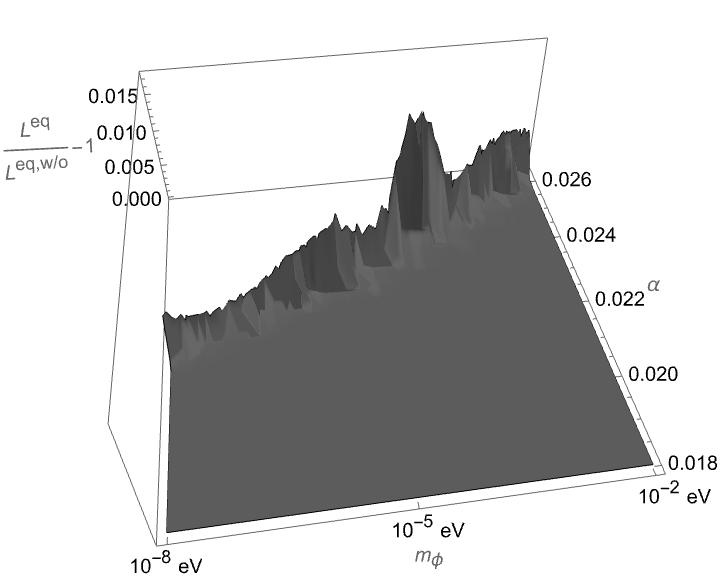}
\caption{Plot of the enhancement of the equilibrium luminosity against $\alpha$ and $m_\phi$. $L^{\textrm{eq}}$ is the equilibrium luminosity, while $L^{\textrm{eq,w/o}}$ is the equilibrium luminosity found in \cite{Rosa_2018}, i.e. ignoring the effects of Schwinger pairs. The same unstable region shown in Fig. 2 is also visible here. In the part of parameter space where an equilibrium state exists, we can see that the equilibrium luminosity is hardly enhanced at all, which means that the role of Schwinger production is negligible to the final state of the BLAST. However, close to the unstable region, there is a region where the enhancement becomes more significant. Note that computational constraints limit the resolution of the section of the plot with the greatest enhancement. This means that, even though the largest enhancement displayed on this plot is $.02$, it is likely that even greater enhancements can be achieved.}
\end{figure}

The buildup of Schwinger pairs suppresses the decay of axions into photons, resulting in a lower equilibrium value for the photon number. The equilibrium value may be found by taking the expression from \cite{Rosa_2018} and substituting $\Gamma_\phi\to\Gamma'_\phi$, which yields the expression
\be {N_\gamma^{\textrm{eq}}}^2\left(1-\frac{8e^2\Gamma_{\textrm{Schw}}^{\textrm{eq}}}{m_em_\phi^2V\Gamma_{e^\pm}^{\textrm{eq}}}\right)={N_\gamma^{\textrm{eq,w/o}}}^2, \ee
where $\Gamma_{\textrm{Schw}}^{\textrm{eq}}$ and $\Gamma_{e^\pm}^{\textrm{eq}}$ represent $\Gamma_{\textrm{Schw}}$ and $\Gamma_{e^\pm}$ evaluated at $N_\gamma^{\textrm{eq}}$, and $N_\gamma^{\textrm{eq,w/o}}=\frac{\Gamma_s}{C_1\Gamma_\phi}$ is the expression for $N_\gamma^{\textrm{eq}}$ calculated without accounting for the Schwinger effect. This equation does not have a solution for all regions of parameter space; for some values of $m_\phi$ and $\alpha$, it is simply impossible for the system to reach equilibrium. This represents the case where the suppression of axion decay is so great that the photons will never build up to the point where they come into equilibrium with the axions. While the number of photons will reach or approach $N_\gamma^{\textrm{ceiling}}$, the number of axions will continue growing unbounded. This is reflected mathematically by the fact that $C_1\Gamma'_\phi N_\gamma$ has some maximum attainable value, and, if this maximum value is less than $\Gamma_s$, then $N_\phi$ will increase without bound. This will eventually result in the black hole spinning down to the point where superradiance ceases. These regions of parameter space where the system is unstable are shown in Fig. 3.

It should be noted that, while this paper has focused on axion decay, which cannot happen when the plasma frequency exceeds $\frac{m_\phi}{2}$, it has been shown that axion clouds may still emit light after exceeding this limit, via axion annihilation, i.e. $\phi+\phi\to\gamma+\gamma$ \cite{Sen_2018,spieksma2023superradianceaxioniccouplingsplasma}. Indeed, it is possible to have any number of axions annihilate to produce two photons, $n\phi\to\gamma+\gamma$, with each individual annihilation process shutting off when $\omega_p>n\frac{m_\phi}{2}$. These higher-order processes will become more relevant in the unstable case, as the axion cloud is expected to reach greater densities than is possible outside of the unstable case, and these processes may in fact allow the system to reach equilibrium (making the term ``unstable case" something of a misnomer; it is only unstable in this particular analysis, which ignores higher-level processes). This, however, is beyond the scope of this paper.

In regions of parameter space where the system is stable, the equilibrium photon number is enhanced by a factor of $\frac{\Gamma_\phi}{\Gamma'_\phi}$. This enhancement is plotted in Fig. 4. It is somewhat counterintuitive that the Schwinger effect, which has the consequence of suppressing axion decay, enhances the equilibrium luminosity of axion lasers; the reason for this enhancement is that, with the Schwinger effect reducing the axion decay rate, a greater number of photons are needed to stimulate axion decay to the point where equilibrium is reached. Thus, when the enhancement is large, we would expect $N_\gamma$ to grow, as a function of time, more slowly than when the enhancement is small; however, it will take a longer time and more photons to reach equilibrium.

For the majority of parameter space (where equilibrium is possible), this enhancement is negligible, as can be seen in Fig. 4. In regions where the enhancement is negligible, the behavior of the BLAST may be modeled without considering the impact of the Schwinger effect. However, we can see that the enhancement to the photon number (and, consequently, the luminosity) increases somewhat as one approaches the region of parameter space where the system becomes unstable. The result is that there is a small region of parameter space, located right next to the region where BLASTs become unstable, where the BLASTS are stable but enhanced. It is in this region where we would expect to find the strongest stable BLASTs.

\section{Conclusion}

When an axion laser, such as a BLAST, becomes sufficiently strong, it can produce an electric field close to the critical Schwinger limit, resulting in the creation of electron-positron pairs. This has three effects on the dynamics of the photon and axion number: photons are lost as their energy is converted into electrons and positrons; electrons and positrons annihilate into photons, but with a higher energy than the photons produced by axion decay; and the buildup of an electron-positron plasma imparts an effective mass on the photon, slowing the rate of axion decay. We found that the third is the dominant phenomenon. How this alters the behavior of a BLAST depends on its parameters, specifically the values of $\alpha$ and $m_\phi$. 

To summarize, the parameter space of a BLAST may be divided into three regions:
\begin{itemize}
\item \textbf{Unenhanced:} In this region, the electron-positron plasma produced by the Schwinger effect has a negligible impact on the dynamics of the BLAST, and the analysis from \cite{Rosa_2018} applies.
\item \textbf{Unstable:} In this region, the system never reaches equilibrium. While the photon number $N_\gamma$ will increase after the onset of lasing, the buildup of an electron-positron plasma suppresses the axion decay rate, so that the axion number $N_\phi$ continues to rise. The annihilation of multiple axions into two photons may eventually bring the system to equilibrium; otherwise, the unbounded growth continues until the black hole spins down, at which point the axion no longer experiences superradiant growth.
\item \textbf{Enhanced:} In between the previous two regions, the BLAST has an equilibrium state, but its equilibrium luminosity is enhanced compared to what is predicted in \cite{Rosa_2018}. The BLAST will take a longer time to reach this equilibrium state, compared to in the unenhanced case.
\end{itemize}

It is important to note that, in the first and third cases, the system will eventually reach its equilibrium state (likely undergoing damped oscillations about said equilibrium state). This is in contrast to what was predicted in \cite{Rosa_2018}, which speculated that the Schwinger effect would ``restart" the system periodically. We therefore should only expect a BLAST to exhibit periodic behavior if we happen to observe it in its initial stages, before it has reached its equilibrium state.

This work has focused on the role of the Schwinger effect, and the resulting electron-positron plasma, in the behavior of BLASTs, but this is not the only means by which a plasma may come to interact with the axion cloud. Black holes are expected to be surrounded by a plasma, acquired via accretion, and the role of this plasma in the behavior of BLASTs is nontrivial \cite{Sen_2018,Bo_kovi__2019,spieksma2023superradianceaxioniccouplingsplasma}. More study is needed to see how the Schwinger-produced electron-positron plasma interacts with this preexisting, accreted plasma.

The equilibrium luminosities we have calculated are quite large in comparison to the mass of a primordial black hole. For example, if $m_\phi=10^{-5}\textrm{ eV}$, then the largest a black hole can be while remaining in the stable region of parameter space is $6.5*10^{28}\textrm{ kg}$, and, if such a black hole were to remain in that equilibrium state continuously, it would evaporate after $7*10^5$ years. In practice, such a black hole would undergo spin-down before this happened. Since the $2p$-state has magnetic quantum number $m=1$, each axion produced by superradiance carries with it angular momentum $\hbar$, and therefore the black hole's spin changes at a rate of $\frac{d\tilde{a}}{dt}=-\frac{\Gamma_sm_\phi}{\alpha M_{BH}}N_\phi$. As the black hole spins down, the superradiant growth of axions slows and eventually stops. Thus, BLASTS represent a potentially important aspect of the spin evolution of rotating black holes.

This paper has focused on the role of the Schwinger effect in one type of axion lasing system, namely, BLASTs. However, there are many other axion lasers that are of research interest (see \cite{Chen:2023jki} for a recent review). It is conceivable that the Schwinger effect may play a significant role in these axion lasers as well. Axion lasers are a promising method by which we might probe the existence of scalar dark matter. Understanding the role of the Schwinger effect in these processes may be necessary in order to develop accurate models of axion lasers.

An important caveat to these results is that our calculation of the rate of pair production ignored the phenomenon of axion assistance to the Schwinger effect \cite{Domcke_2021,Domcke_2022}. Because a BLAST occurs in an axion field, we should expect the rate of pair production to be greater than what is calculated using the standard Schwinger effect (i.e., based only on the electromagnetic field, without considering the axion field). The rate of Schwinger production used in this paper therefore serves as a lower bound. The closeness this lower bound to reality will depend on the axion-electron coupling, $g_{ae}$, with the true rate of pair production approaching the rate used in this paper only in the limit $g_{ae}\to0$. We may predict that increasing $g_{ae}$ will have the effect of expanding the unbounded region of parameter space, as it becomes easier for the Schwinger effect to create a plasma of sufficient density that axion decay is shut off. However, future research is needed to examine, in greater detail, the behavior of BLASTs in the case of $g_{ae}>0$.

It is also worth noting that we have assumed that all particles are uniformly distributed throughout the $2p$-cloud; however, the simulations conducted in \cite{spieksma2023superradianceaxioniccouplingsplasma} have demonstrated that this approximation can fail to predict significant phenomena. It is therefore necessary to do away with this assumption as well, which will entail replacing the analytical calculations in this paper with numerical simulations.

\section*{Acknowledgements}

The author thanks Devin Walker and João Rosa for their valuable feedback, as well as Roberto Onofrio and Jens Mahlmann for helpful conversations. This work was supported by the National Science Foundation under grant OIA-2033382. 

The data that support the findings of this article are openly available \cite{data}.

\appendix
\section{Calculation of the rate of pair production}\label{gammacalc}

In this appendix we perform the integral in eq.~\ref{Gammaschwdef} in order to attain an expression for $\Gamma_{\textrm{Schw}}$. We begin by rewriting our integral in terms only of $E_\parallel$, $B_\parallel$, and $\theta$:\begin{widetext}
\begin{align}
\Gamma_{\textrm{Schw}}=\frac{e^2V}{4\pi^3\sigma_{\textrm{EM}}^6}\int_0^\pi\int_0^\infty\int_0^\infty\d\theta&\d E_\parallel\d B_\parallel E_\parallel^3B_\parallel^3\left(E_\parallel^2+B_\parallel^2\right) \nonumber\\
&\frac{\sin\theta\lvert\sec^3\theta\rvert e^{-\frac{\sqrt{\left(E_\parallel^2-B_\parallel^2\right)^2+4E_\parallel^2B_\parallel^2\sec^2\theta}}{2\sigma_{\textrm{EM}}^2}}}{\sqrt{\left(E_\parallel^2-B_\parallel^2\right)^2+4E_\parallel^2B_\parallel^2\sec^2\theta}}\sum_{n=1}^\infty\frac{1}{n}\coth\frac{n\pi B_\parallel}{E_\parallel}e^{-n\pi\frac{E_c}{E_\parallel}}.
\end{align}
The integral over $\theta$ may be performed through the substitution $u=\sqrt{\left(E_\parallel^2-B_\parallel^2\right)^2+4E_\parallel^2B_\parallel^2\sec^2\theta}$, yielding
\begin{align}
\label{truegammaschw}\Gamma_{\textrm{Schw}}=\frac{e^2V}{4\pi^3\sigma_{\textrm{EM}}^4}\int_0^\infty\int_0^\infty\d E_\parallel\d B_\parallel E_\parallel B_\parallel\left(E_\parallel^2+B_\parallel^2\right)e^{-\frac{E_\parallel^2+B_\parallel^2}{2\sigma_{\textrm{EM}}^2}}\sum_{n=1}^\infty\frac{1}{n}\coth\frac{n\pi B_\parallel}{E_\parallel}e^{-n\pi\frac{E_c}{E_\parallel}}.
\end{align}
We may expand $\coth$ as a series of exponentials, allowing us to rewrite this as
\begin{align}
\Gamma_{\textrm{Schw}}=\frac{e^2V}{4\pi^3\sigma_{\textrm{EM}}^4}\sum_{n=1}^\infty\frac{1}{n}\int_0^\infty\d E_\parallel E_\parallel e^{-\frac{E_\parallel^2}{2\sigma_{\textrm{EM}}^2}}e^{-n\pi\frac{E_c}{E_\parallel}}\sum_{k=0}^\infty\left(2-\delta_{k0}\right)\int_0^\infty\d B_\parallel B_\parallel\left(E_\parallel^2+B_\parallel^2\right)e^{-\frac{B_\parallel^2}{2\sigma_{\textrm{EM}}^2}}e^{-2nk\pi\frac{B_\parallel}{E_\parallel}}.
\end{align}
To perform the inner integral, we may rewrite the $B_\parallel^2$ exponential as a Meijer G-function, $\exp\left(-\frac{B_\parallel^2}{2\sigma_{\textrm{EM}}^2}\right)=G^{1,0}_{0,1}\left(\frac{B_\parallel^2}{2\sigma_{\textrm{EM}}^2}\left\lvert\begin{tabular}{cc}$-$\\$0$\end{tabular}\right.\right)$, and then make use of the identity \cite{Prudnikov_1989}
\begin{align}
\int_0^\infty \d xx^{\alpha-1}e^{-\sigma x}G^{mn}_{pq}\left(\omega x^\ell\left\lvert\begin{tabular}{c}$(a_p)$\\$(b_q)$\end{tabular}\right.\right)=\frac{\ell^{\alpha-\frac{1}{2}}\sigma^{-\alpha}}{(2\pi)^{\frac{1}{2}(\ell-1)}}G^{m,n+\ell}_{p+\ell,q}\left(\frac{\omega}{(\sigma/\ell)^\ell}\left\lvert\begin{tabular}{c}$\frac{1-\alpha}{\ell},\frac{2-\alpha}{\ell},\ldots,\frac{\ell-\alpha}{\ell},(a_p)$\\$(b_q)$\end{tabular}\right.\right).\label{meijeridentity}
\end{align}
This identity requires a number of conditions for its validity, but the only ones relevant to this discussion are that $\sigma$ and $\omega$ must both be nonzero; consequently, we must separate the $k=0$ term from the $k\neq0$ terms, applying the above identity only to the latter. This yields
\begin{align}
\Gamma_{\textrm{Schw}}&=\frac{e^2V}{4\pi^3\sigma_{\textrm{EM}}^4}\sum_{n=1}^\infty\frac{1}{n}\int_0^\infty\d E_\parallel E_\parallel e^{-\frac{E_\parallel^2}{2\sigma_{\textrm{EM}}^2}}e^{-n\pi\frac{E_c}{E_\parallel}}\Bigg(2\sigma_{\textrm{EM}}^4+\sigma_{\textrm{EM}}^2E_\parallel^2 \nonumber\\
&\hspace{1.6cm}+\sum_{k=1}^\infty\frac{E_\parallel^4}{n^4k^4\pi^{4.5}}\left(n^2k^2\pi^2G^{1,2}_{2,1}\left(\frac{E_\parallel^2}{2n^2k^2\pi^2\sigma_{\textrm{EM}}^2}\left\lvert\begin{tabular}{c}$0,-\frac{1}{2}$\\$0$\end{tabular}\right.\right)+G^{1,2}_{2,1}\left(\frac{E_\parallel^2}{2n^2k^2\pi^2\sigma_{\textrm{EM}}^2}\left\lvert\begin{tabular}{c}$-1,-\frac{3}{2}$\\$0$\end{tabular}\right.\right)\right)\Bigg).
\end{align}
Defining $v=E_\parallel^2$:
\begin{align}
\Gamma_{\textrm{Schw}}&=\frac{e^2V}{8\pi^{3.5}\sigma_{\textrm{EM}}^4}\sum_{n=1}^\infty\frac{1}{n}\int_0^\infty\d v e^{-\frac{v}{2\sigma_{\textrm{EM}}^2}}G^{0,2}_{2,0}\left(\frac{4v}{n^2\pi^2E_c^2}\left\lvert\begin{tabular}{cc}$1,\frac{1}{2}$\\$-$\end{tabular}\right.\right)\Bigg(2\sigma_{\textrm{EM}}^4+\sigma_{\textrm{EM}}^2v \nonumber\\
&\hspace{1.6cm}+\sum_{k=1}^\infty\frac{v^2}{n^4k^4\pi^{4.5}}\left(n^2k^2\pi^2G^{1,2}_{2,1}\left(\frac{v}{2n^2k^2\pi^2\sigma_{\textrm{EM}}^2}\left\lvert\begin{tabular}{c}$0,-\frac{1}{2}$\\$0$\end{tabular}\right.\right)+G^{1,2}_{2,1}\left(\frac{v}{2n^2k^2\pi^2\sigma_{\textrm{EM}}^2}\left\lvert\begin{tabular}{c}$-1,-\frac{3}{2}$\\$0$\end{tabular}\right.\right)\right)\Bigg).
\end{align}
The first two terms may be integrated using eq.~\ref{meijeridentity}. Integrating the higher-order terms requires the use of the bivariate Meijer G-function \cite{agarwal1965extension,sharma1965generalised,hai1992}, defined as
\begin{align}
G^{m_1,n_1:m_2,n_2;m_3,n_3}_{\,p_1,\,q_1\,:\,p_2,\,q_2\,;\,p_3,\,q_3}\left(x,y\left\lvert\begin{tabular}{ccccc}$\left(\alpha^{(1)}_{p_1}\right)$&:&$\left(\alpha^{(2)}_{p_2}\right)$&;&$\left(\alpha^{(3)}_{p_3}\right)$\\$\left(\beta^{(1)}_{q_1}\right)$&:&$\left(\beta^{(2)}_{q_2}\right)$&;&$\left(\beta^{(3)}_{q_3}\right)$\end{tabular}\right.\right)=\frac{-1}{4\pi^2}\int_{L_1}\int_{L_2}\Psi_1(s+t)\Psi_2(s)\Psi_3(t)x^sy^t\d s\d t,
\end{align}
where $L_1$ and $L_2$ are suitable contours and
\be
\Psi_i(u)=\frac{\displaystyle\prod_{j=1}^{m_i}\Gamma\left(\beta^{(i)}_j-u\right)\prod_{j=1}^{n_i}\Gamma\left(1-\alpha^{(i)}_j+u\right)}{\displaystyle\prod_{j=m_i+1}^{q_i}\Gamma\left(1-\beta^{(i)}_j+u\right)\prod_{j=n_i+1}^{p_i}\Gamma\left(\alpha^{(i)}_j-u\right)}.
\ee
When $p_1=q_1=0$, the bivariate Meijer G-function becomes a product of two univariate Meijer G-functions. The integral may then be performed using the identity \cite{Shah1973}
\begin{align}
&\int_0^\infty \d xx^{\lambda-1}e^{-\mu x}G^{m_1,n_1:m_2,n_2;m_3,n_3}_{\,p_1\,,q_1\,:\,p_2,\,q_2\,;\,p_3,\,q_3}\left(\rho x,\sigma x\left\lvert\begin{tabular}{ccccc}$\left(\alpha^{(1)}_{p_1}\right)$&:&$\left(\alpha^{(2)}_{p_2}\right)$&;&$\left(\alpha^{(3)}_{p_3}\right)$\\$\left(\beta^{(1)}_{q_1}\right)$&:&$\left(\beta^{(2)}_{q_2}\right)$&;&$\left(\beta^{(3)}_{q_3}\right)$\end{tabular}\right.\right) \nonumber\\
&\hspace{5cm}=\mu^{-\lambda}G^{m_1,n_1+1:m_2,n_2;m_3,n_3}_{\,p_1\,,\,\,\,\,\,\,q_1\,\,\,:\,p_2,\,q_2\,;\,p_3,\,q_3}\left(\frac{\rho}{\mu},\frac{\sigma}{\mu}\left\lvert\begin{tabular}{ccccc}$1-\lambda,\left(\alpha^{(1)}_{p_1}\right)$&:&$\left(\alpha^{(2)}_{p_2}\right)$&;&$\left(\alpha^{(3)}_{p_3}\right)$\\$\left(\beta^{(1)}_{q_1}\right)$&:&$\left(\beta^{(2)}_{q_2}\right)$&;&$\left(\beta^{(3)}_{q_3}\right)$\end{tabular}\right.\right),
\end{align}
yielding
\begin{align}
\Gamma_{\textrm{Schw}}&=\sum_{n=1}^\infty\frac{e^2V\sigma_{\textrm{EM}}^2}{2\pi^{3.5}n}\left(G^{0,3}_{3,0}\left(\frac{8\sigma_{\textrm{EM}}^2}{n^2\pi^2E_c^2}\left\lvert\begin{tabular}{c}$1,\frac{1}{2},0$\\$-$\end{tabular}\right.\right)+G^{0,3}_{3,0}\left(\frac{8\sigma_{\textrm{EM}}^2}{n^2\pi^2E_c^2}\left\lvert\begin{tabular}{c}$1,\frac{1}{2},-1$\\$-$\end{tabular}\right.\right)\right) \nonumber\\
&\hspace{.5cm}+\sum_{n=1}^\infty\sum_{k=1}^\infty\frac{e^2V\sigma_{\textrm{EM}}^2}{n^5k^4\pi^{8}}\left(n^2k^2\pi^2G^{0,1:0,2;1,2}_{1,0:2,0;2,1}\left(\frac{8\sigma_{\textrm{EM}}^2}{n^2\pi^2E_c^2},\frac{1}{n^2k^2\pi^2}\left\lvert\begin{tabular}{ccccc}$-2$&$:$&$1,\frac{1}{2}$&$;$&$0,-\frac{1}{2}$\\$-$&$:$&$-$&$;$&$0$\end{tabular}\right.\right)\right. \nonumber\\
&\hspace{5cm}\left.+G^{0,1:0,2;1,2}_{1,0:2,0;2,1}\left(\frac{8\sigma_{\textrm{EM}}^2}{n^2\pi^2E_c^2},\frac{1}{n^2k^2\pi^2}\left\lvert\begin{tabular}{ccccc}$-2$&$:$&$1,\frac{1}{2}$&$;$&$-1,-\frac{3}{2}$\\$-$&$:$&$-$&$;$&$0$\end{tabular}\right.\right)\right).
\end{align}
With some manipulation, we may rewrite this as
\begin{align}
\Gamma_{\textrm{Schw}}=\frac{e^2V\sigma_{\textrm{EM}}^2}{2\pi^{3.5}}\sum_{n=1}^\infty\frac{1}{n}\Bigg(&G^{0,4}_{4,1}\left(\frac{8\sigma_{\textrm{EM}}^2}{n^2\pi^2E_c^2}\left\lvert\begin{tabular}{c}$1,\frac{1}{2},0,-2$\\$-1$\end{tabular}\right.\right) \nonumber\\
&\hspace{.5cm}+\frac{2}{\sqrt{\pi}}\sum_{k=1}^\infty G^{0,1:0,3;1,2}_{1,0:3,1;2,1}\left(\frac{8\sigma_{\textrm{EM}}^2}{n^2\pi^2E_c^2},\frac{1}{n^2k^2\pi^2}\left\lvert\begin{tabular}{ccccc}$0$&$:$&$1,\frac{1}{2},-2$&$;$&$1,\frac{1}{2}$\\$-$&$:$&$-1$&$;$&$1$\end{tabular}\right.\right)\Bigg).
\end{align}

We now wish to evaluate this in the limit $\sigma_{\textrm{EM}}\ll E_c$. To do this, we first rewrite the bivariate Meijer G-function as
\begin{align}
&G^{0,1:0,3;1,2}_{1,0:3,1;2,1}\left(\frac{8\sigma_{\textrm{EM}}^2}{n^2\pi^2E_c^2},\frac{1}{n^2k^2\pi^2}\left\lvert\begin{tabular}{ccccc}$0$&$:$&$1,\frac{1}{2},-2$&$;$&$1,\frac{1}{2}$\\$-$&$:$&$-1$&$;$&$1$\end{tabular}\right.\right) \nonumber\\
&\hspace{3cm}=\frac{1}{2\pi i}\int_L\Gamma(t)\Gamma\left(\frac{1}{2}+t\right)\Gamma(1-t)G^{0,4}_{4,1}\left(\frac{8\sigma_{\textrm{EM}}^2}{n^2\pi^2E_c^2}\left\lvert\begin{tabular}{c}$1,\frac{1}{2},-2,-t$\\$-1$\end{tabular}\right.\right)\left(\frac{1}{n^2k^2\pi^2}\right)^t\d t.
\end{align}
In the aforementioned limit,
\be G^{0,4}_{4,1}\left(\frac{8\sigma_{\textrm{EM}}^2}{n^2\pi^2E_c^2}\left\lvert\begin{tabular}{c}$1,\frac{1}{2},-2,-t$\\$-1$\end{tabular}\right.\right)\approx\frac{2\pi}{\sqrt{3}}e^{-3\left(\frac{8\sigma_{\textrm{EM}}^2}{n^2\pi^2E_c^2}\right)^{\frac{-1}{3}}}\left(\frac{8\sigma_{\textrm{EM}}^2}{n^2\pi^2E_c^2}\right)^{-\frac{1}{2}-\frac{t}{3}}, \ee
and thus we get
\begin{align}
&G^{0,1:0,3;1,2}_{1,0:3,1;2,1}\left(\frac{8\sigma_{\textrm{EM}}^2}{n^2\pi^2E_c^2},\frac{1}{n^2k^2\pi^2}\left\lvert\begin{tabular}{ccccc}$0$&$:$&$1,\frac{1}{2},-2$&$;$&$1,\frac{1}{2}$\\$-$&$:$&$-1$&$;$&$1$\end{tabular}\right.\right) \nonumber\\
&\hspace{3cm}\approx\frac{n\pi^2E_c}{\sqrt{6}\sigma_{\textrm{EM}}}e^{-3\left(\frac{8\sigma_{\textrm{EM}}^2}{n^2\pi^2E_c^2}\right)^{\frac{-1}{3}}}\frac{1}{2\pi i}\int_L\Gamma(t)\Gamma\left(\frac{1}{2}+t\right)\Gamma(1-t)\left(\frac{1}{k^2}\left(\frac{E_c^2}{8n^4\pi^4\sigma_{\textrm{EM}}^2}\right)^{\frac{1}{3}}\right)^t\d t.
\end{align}
The sum over $k$ becomes a Riemann zeta function:
\begin{align}
&\sum_{k=1}^\infty G^{0,1:0,3;1,2}_{1,0:3,1;2,1}\left(\frac{8\sigma_{\textrm{EM}}^2}{n^2\pi^2E_c^2},\frac{1}{n^2k^2\pi^2}\left\lvert\begin{tabular}{ccccc}$0$&$:$&$1,\frac{1}{2},-2$&$;$&$1,\frac{1}{2}$\\$-$&$:$&$-1$&$;$&$1$\end{tabular}\right.\right) \nonumber\\
&\hspace{1cm}=\frac{n\pi^2E_c}{\sqrt{6}\sigma_{\textrm{EM}}}e^{-3\left(\frac{8\sigma_{\textrm{EM}}^2}{n^2\pi^2E_c^2}\right)^{\frac{-1}{3}}}\frac{1}{2\pi i}\int_L\Gamma(t)\Gamma\left(\frac{1}{2}+t\right)\Gamma(1-t)\left(\left(\frac{E_c^2}{8n^4\pi^4\sigma_{\textrm{EM}}^2}\right)^{\frac{1}{3}}\right)^t\zeta(2t)\d t \nonumber\\
&\hspace{1cm}=\frac{n\pi^2E_c}{\sqrt{6}\sigma_{\textrm{EM}}}e^{-3\left(\frac{8\sigma_{\textrm{EM}}^2}{n^2\pi^2E_c^2}\right)^{\frac{-1}{3}}}\frac{1}{2\pi i}\int_0^\infty\frac{x^{-1}\d x}{e^x-1}\int_L\frac{\Gamma(t)\Gamma\left(\frac{1}{2}+t\right)}{\Gamma(2t)}\Gamma(1-t)\left(x^2\left(\frac{E_c^2}{8n^4\pi^4\sigma_{\textrm{EM}}^2}\right)^{\frac{1}{3}}\right)^t\d t,
\end{align}
where we have used the zeta function's integral representation,
\be \zeta(t)=\sum_{k=1}^\infty\frac{1}{k^t}=\frac{1}{\Gamma(t)}\int_0^\infty\frac{x^{t-1}}{e^x-1}\d x. \ee
With some manipulation, the inner integral may be rewritten as a Meijer G-function with a known form, yielding
\begin{align}
&\sum_{k=1}^\infty G^{0,1:0,3;1,2}_{1,0:3,1;2,1}\left(\frac{8\sigma_{\textrm{EM}}^2}{n^2\pi^2E_c^2},\frac{1}{n^2k^2\pi^2}\left\lvert\begin{tabular}{ccccc}$0$&$:$&$1,\frac{1}{2},-2$&$;$&$1,\frac{1}{2}$\\$-$&$:$&$-1$&$;$&$1$\end{tabular}\right.\right) \nonumber\\
&\hspace{1cm}=\frac{n\pi^2E_c}{\sqrt{6}\sigma_{\textrm{EM}}}e^{-3\left(\frac{8\sigma_{\textrm{EM}}^2}{n^2\pi^2E_c^2}\right)^{\frac{-1}{3}}}\left(\frac{E_c^2}{8n^4\pi^4\sigma_{\textrm{EM}}^2}\right)^{\frac{1}{3}}\frac{\sqrt{\pi}}{2}\int_0^\infty\frac{x\d x}{e^x-1}e^{-\frac{x^2}{4}\left(\frac{E_c^2}{8n^4\pi^4\sigma_{\textrm{EM}}^2}\right)^{\frac{1}{3}}}.
\end{align}
Our expression for the total rate of pair production then becomes
\begin{align}
\Gamma_{\textrm{Schw}}=\frac{e^2E_cV\sigma_{\textrm{EM}}}{2\sqrt{6}\pi^{1.5}}\sum_{n=1}^\infty e^{-3\left(\frac{8\sigma_{\textrm{EM}}^2}{n^2\pi^2E_c^2}\right)^{\frac{-1}{3}}}\left(1+\left(\frac{E_c^2}{8n^4\pi^4\sigma_{\textrm{EM}}^2}\right)^{\frac{1}{3}}\int_0^\infty\frac{x\d x}{e^x-1}e^{-\frac{x^2}{4}\left(\frac{E_c^2}{8n^4\pi^4\sigma_{\textrm{EM}}^2}\right)^{\frac{1}{3}}}\right).
\end{align}
Terms in this summation vanish as $n$ increases, so that we may assume $n$ is small, and therefore $\frac{E_c^2}{8n^4\pi^4\sigma_{\textrm{EM}}^2}\gg1$. In this limit,
\be \left(\frac{E_c^2}{8n^4\pi^4\sigma_{\textrm{EM}}^2}\right)^{\frac{1}{3}}\int_0^\infty\frac{x\d x}{e^x-1}e^{-\frac{x^2}{4}\left(\frac{E_c^2}{8n^4\pi^4\sigma_{\textrm{EM}}^2}\right)^{\frac{1}{3}}}\approx\sqrt{\pi}\left(\frac{E_c^2}{8n^4\pi^4\sigma_{\textrm{EM}}^2}\right)^{\frac{1}{6}}-1, \ee\end{widetext}
and so we get
\begin{align}
\Gamma_{\textrm{Schw}}&\approx\frac{e^2E_c^2V}{8\sqrt{3}\pi}\sum_{n=1}^\infty e^{-3\left(\frac{8\sigma_{\textrm{EM}}^2}{n^2\pi^2E_c^2}\right)^{\frac{-1}{3}}}\left(\frac{8\sigma_{\textrm{EM}}^2}{n^2\pi^2E_c^2}\right)^{\frac{1}{3}}.
\end{align}
Substituting the expression for $\sigma_{\textrm{EM}}$ found from eq.~\ref{sigmafind}, and using the definition for $N_\gamma^{\textrm{Schw}}$, yields eq.~\ref{Schwrate}. Numerical calculations show that eq.~\ref{Schwrate} approximates eq.~\ref{truegammaschw} well for $N_\gamma\lesssim.1N_\gamma^{\textrm{Schw}}$, and continues to be accurate to within an order of magnitude for all $N_\gamma<N_\gamma^{\textrm{Schw}}$.

\section{Behavior of functions of $\chi$}\label{chianalysis}

The dimensionless parameter $\chi$ represents the rate at which electrons and positrons are accelerated to relativistic velocities by radiation pressure. A number of quantities may be expressed solely as a function of $\chi$, and in this appendix we establish the mathematical tools to analyze these functions.

It is helpful to rescale quantities of length by a factor of $\frac{\chi}{\Delta r}$; we label quantities that have been rescaled in this way with a hat, e.g. $\hat{\rho}_0=\chi\frac{\rho_0}{\Delta r}$. With this notation, the relation between the location of an electron or positron's creation and the time it takes for that electron or positron to exit the $2p$-cloud may be written as
\begin{widetext}\be
\chi-\hat{\rho}_0=\sqrt{\hat{T}_{e^\pm}\left(2+\hat{T}_{e^\pm}\right)}+\frac{\pi}{3}-2\tan^{-1}\frac{\sqrt{3}\left(1+\hat{T}_{e^\pm}\right)+2\sqrt{\hat{T}_{e^\pm}\left(2+\hat{T}_{e^\pm}\right)}}{3+\hat{T}_{e^\pm}}\label{app1}, \ee\end{widetext}
and the escape rate is
\be \Gamma_{e^\pm}=\frac{\chi^3}{6\int_0^\chi \hat{T}_{e^\pm}\hat{\rho}_0\d\hat{\rho}_0}\Gamma_\gamma. \ee
This makes it clear that $\frac{\Gamma_{e^\pm}}{\Gamma_\gamma}$ is a function only of $\chi$. Its asymptotic behavior may be derived by noting that, in the limit $\chi\to\infty$, the arctangent term in eq.~\ref{app1} becomes insignificant, and therefore

\be \Gamma_{e^\pm}\xrightarrow[\chi\to\infty]{}\frac{\chi^3}{6\int_0^\chi\left(\sqrt{\left(\chi-\hat{\rho}_0\right)^2+1}-1\right)\hat{\rho}_0\d\hat{\rho}_0}\Gamma_\gamma. \ee
This may be evaluated to find that
\be \lim_{\chi\to\infty}\Gamma_{e^\pm}=\Gamma_\gamma, \ee
which is what one would expect physically.

An integral which appears in pair annihilation, when written using hat notation, is
\begin{widetext}\be \Gamma_{\textrm{ann}}V=\frac{8}{\chi^2}\int_0^\chi\d\hat{\rho}\int_0^{\hat{\rho}}\d\hat{\rho}_{0e^+}\int_0^{\hat{\rho}}\d\hat{\rho}_{0e^-}\frac{\hat{\rho}_{0e^+}\hat{\rho}_{0e^-}}{\hat{\rho}^3}\lvert v_{e^+}-v_{e^-}\rvert\sigma(v_{\textrm{com}}), \ee
where
\be \hat{\rho}-\hat{\rho}_{0e^\pm}=\frac{v_{e^\pm}}{\sqrt{1-v_{e^\pm}^2}}-\tan^{-1}\frac{v_{e^\pm}}{\sqrt{1-v_{e^\pm}^2}}. \ee
The asymptotic behavior of $\Gamma_{\textrm{ann}}V$ may be examined in a similar manner to $\Gamma_{e^\pm}$, although in this case it is convenient to examine $\frac{\d\left(\chi^2\Gamma_{\textrm{ann}}V\right)}{\d\chi}$:
\be \frac{\d\left(\chi^2\Gamma_{\textrm{ann}}V\right)}{\d\chi}\xrightarrow[\chi\to\infty]{}\frac{8}{\chi^3}\int_0^{\frac{\chi}{\sqrt{\chi^2+1}}}\int_0^{\frac{\chi}{\sqrt{\chi^2+1}}}\d v_{e^+}\d v_{e^-}\gamma_{e^+}^3\gamma_{e^-}^3\left(\chi-\frac{v_{e^+}}{\sqrt{1-v_{e^+}^2}}\right)\left(\chi-\frac{v_{e^-}}{\sqrt{1-v_{e^-}^2}}\right)\lvert v_{e^+}-v_{e^-}\rvert\sigma(v_{\textrm{com}}). \ee
We may convert the double integral into a single integral by differentiating three times:
\be \frac{\d^3}{\d\chi^3}\chi^3\frac{\d\left(\chi^2\Gamma_{\textrm{ann}}V\right)}{\d\chi}\xrightarrow[\chi\to\infty]{}16\int_0^{\frac{\chi}{\sqrt{\chi^2+1}}}\d v_{e^-}\gamma_{e^-}^3\left(3+\left(\chi-\frac{v_{e^-}}{\sqrt{1-v_{e^-}^2}}\right)\frac{\d}{\d\chi}\right)\left(\lvert v_{e^+}-v_{e^-}\rvert\sigma(v_{\textrm{com}})\bigg\rvert_{v_{e^+}=\frac{\chi}{\sqrt{\chi^2+1}}}\right) \ee
In the limit $\chi\to\infty$, the integrand becomes
\be \gamma_{e^-}^3\left(3+\left(\chi-\frac{v_{e^-}}{\sqrt{1-v_{e^-}^2}}\right)\frac{\d}{\d\chi}\right)\left(\lvert v_{e^+}-v_{e^-}\rvert\sigma(v_{\textrm{com}})\bigg\rvert_{v_{e^+}=\frac{\chi}{\sqrt{\chi^2+1}}}\right)\xrightarrow[\chi\to\infty]{}\frac{3(1+v_{e^-})\ln\left(2\chi\sqrt{\frac{1-v_{e^-}}{1+v_{e^-}}}\right)}{8\chi^2\left(1-v_{e^-}^2\right)^{\frac{3}{2}}}\sigma_T. \ee
When this is integrated, we arrive at our asymptotic form for $\Gamma_{\textrm{ann}}V$:
\be \Gamma_{\textrm{ann}}V\xrightarrow[\chi\to\infty]{}\frac{3(\ln\chi)^2}{2\chi^2}\sigma_T. \ee
eq.~\ref{gasymptote} follows straightforwardly.\end{widetext}

\bibliography{\jobname}

\end{document}